\def\sphinxdocclass{report}
\documentclass[letterpaper,10pt,english]{sphinxmanual}
\ifdefined\pdfpxdimen
   \let\sphinxpxdimen\pdfpxdimen\else\newdimen\sphinxpxdimen
\fi \sphinxpxdimen=.75bp\relax

\PassOptionsToPackage{warn}{textcomp}
\usepackage[utf8]{inputenc}
\ifdefined\DeclareUnicodeCharacter
  \ifdefined\DeclareUnicodeCharacterAsOptional
    \def\sphinxDUC#1{\DeclareUnicodeCharacter{"#1}}
  \else
    \let\sphinxDUC\DeclareUnicodeCharacter
  \fi
  \sphinxDUC{00A0}{\nobreakspace}
  \sphinxDUC{2500}{\sphinxunichar{2500}}
  \sphinxDUC{2502}{\sphinxunichar{2502}}
  \sphinxDUC{2514}{\sphinxunichar{2514}}
  \sphinxDUC{251C}{\sphinxunichar{251C}}
  \sphinxDUC{2572}{\textbackslash}
\fi
\usepackage{cmap}
\usepackage[T1]{fontenc}
\usepackage{amsmath,amssymb,amstext}
\usepackage{babel}

\usepackage{times}
\expandafter\ifx\csname T@LGR\endcsname\relax
\else
  \substitutefont{LGR}{\rmdefault}{cmr}
  \substitutefont{LGR}{\sfdefault}{cmss}
  \substitutefont{LGR}{\ttdefault}{cmtt}
\fi
\expandafter\ifx\csname T@X2\endcsname\relax
  \expandafter\ifx\csname T@T2A\endcsname\relax
  \else
    \substitutefont{T2A}{\rmdefault}{cmr}
    \substitutefont{T2A}{\sfdefault}{cmss}
    \substitutefont{T2A}{\ttdefault}{cmtt}
  \fi
\else
  \substitutefont{X2}{\rmdefault}{cmr}
  \substitutefont{X2}{\sfdefault}{cmss}
  \substitutefont{X2}{\ttdefault}{cmtt}
\fi

\usepackage[Bjarne]{fncychap}
\usepackage{sphinx}

\fvset{fontsize=\small}
\usepackage{geometry}

\usepackage{hyperref}
\usepackage{hypcap}% it must be loaded after hyperref.
\addto\captionsenglish{}

\usepackage{sphinxmessages}
\title{Conjure Documentation}
\date{Oct 01, 2019}
\release{2.3.0}
\author{Özgür Akgün, András Salamon}
\newcommand{\sphinxlogo}{\vbox{}}
\renewcommand{\releasename}{Release}
\makeindex
\begin{document}

\pagestyle{empty}
\sphinxmaketitle
\pagestyle{plain}
\sphinxtableofcontents
\pagestyle{normal}
\phantomsection\label{\detokenize{index::doc}}

\chapter{Welcome}
\label{\detokenize{welcome:welcome}}\label{\detokenize{welcome:id1}}\label{\detokenize{welcome::doc}}
Welcome to the documentation of Conjure!

Conjure is an automated modelling tool for Constraint Programming.

In this documentation, you will find the following.
\begin{itemize}
\item {} 
A brief introduction to Conjure,

\item {} 
installation instructions,

\item {} 
a description of how to use Conjure through its command line user interface,

\item {} 
a list of Conjure’s features,

\item {} 
a description of Conjure’s input language Essence, and

\item {} 
a collection of simple demonstrations of Conjure’s use.

\end{itemize}

\chapter{Introduction}
\label{\detokenize{introduction:introduction}}\label{\detokenize{introduction:id1}}\label{\detokenize{introduction::doc}}
Conjure is an automated constraint modelling tool for Constraint Programming.

Its input language, Essence, is a high level problem specification language.
Essence allows writing problem specifications at a high level of abstraction and without having to make a lot of low level modelling decisions.

Conjure reads in abstract problem specifications (in Essence) and produces concrete constraint programming models (in Essence’).
Essence’ is a solver independent constraint modelling language.
Using the Savile Row tool, an Essence’ model can be instantiated with parameter values and solved using one of several backends.
More information on Savile Row can be found on \sphinxhref{http://savilerow.cs.st-andrews.ac.uk}{its website}.

Conjure works at the problem class level.
A problem class is a parameterised specification of a problem; it does not encode a single problem but a class of problems.
For example, a problem specification for the game of Sudoku is typically parameterised over the hints (the prefilled cells).
A problem specification (or model) at the class level is said to be \sphinxstyleemphasis{instantiated} when values are provided for its parameters.
In the case of a Sudoku, the parameter values are the contents of the hint cells.

Operating at the class level has one very important benefit: Conjure needs to be executed only once to create one (or more) Essence’ models for a problem.
Once the models are generated, they can be used to solve many instances of the same class.

\chapter{Installation}
\label{\detokenize{installation:installation}}\label{\detokenize{installation:id1}}\label{\detokenize{installation::doc}}
Conjure can be installed either by downloading a binary distribution, or by compiling it from source code.

\section{Downloading a binary}
\label{\detokenize{installation:downloading-a-binary}}
Conjure is available as an executable binary for Linux and MacOS.
If it is available for your platform, you can just \sphinxhref{https://www.github.com/conjure-cp/conjure/releases/latest}{download it} and run it.
It may be useful to save the binary under a directory that is in your search PATH, so you do not have to type the full path to the Conjure executable to run it.

For Windows, please use the Linux binaries with the
\sphinxhref{https://en.wikipedia.org/wiki/Windows\_Subsystem\_for\_Linux}{Windows Subsystem for Linux}.

\section{Compiling from source}
\label{\detokenize{installation:compiling-from-source}}
In order to compile Conjure on your computer, please download the source code from \sphinxhref{https://github.com/conjure-cp/conjure}{GitHub}.

\begin{sphinxVerbatim}[commandchars=\\\{\}]
git clone git@github.com:conjure\PYGZhy{}cp/conjure.git
\PYG{n+nb}{cd} conjure
\PYG{n+nv}{BIN\PYGZus{}DIR}\PYG{o}{=}/somewhere/in/your/path make install
\end{sphinxVerbatim}

Conjure is implemented in Haskell, it can be compiled using either \sphinxhref{http://wiki.haskell.org/Cabal-Install}{cabal-install} or \sphinxhref{https://docs.haskellstack.org/en/stable/README/}{stack}.

It comes with a Makefile which will use Stack by default.
The default target in the Makefile will install Stack using the standard procedures (which involves downloading and running a script).
For more precise control, you might want to consider installing the Haskell tools beforehand instead of using the Makefile.

Installation is known to work with
\sphinxhref{http://www.haskell.org/ghc/download\_ghc\_8\_0\_2.html}{GHC-8.0.2},
\sphinxhref{http://www.haskell.org/ghc/download\_ghc\_8\_2\_2.html}{GHC-8.2.2},
\sphinxhref{http://www.haskell.org/ghc/download\_ghc\_8\_4\_4.html}{GHC-8.4.2}, and
\sphinxhref{http://www.haskell.org/ghc/download\_ghc\_8\_6\_5.html}{GHC-8.4.2}.

In addition, a number of supported backend solvers can be compiled using the \sphinxtitleref{make solvers} target.
This target also takes a BIN\_DIR environment variable to control the location of the solver executables.

\section{Installing Savile Row}
\label{\detokenize{installation:installing-savile-row}}
Since Conjure works by generating an Essence’ model, Savile Row is a vital tool when using it.
Savile Row can be downloaded from \sphinxhref{http://savilerow.cs.st-andrews.ac.uk}{its website}.

You do not need to download Savile Row separately when you compile Conjure from source.
An up-to-date version of Savile Row is also copied next to the Conjure executable.

\chapter{Command Line Interface}
\label{\detokenize{cli:command-line-interface}}\label{\detokenize{cli:cli}}\label{\detokenize{cli::doc}}
Conjure supports a number of commands.
A command is provided as the first argument to Conjure on the command line.
It is followed by a number of mandatory arguments (if any) depending on the command, and a number of optional arguments.

Some command line arguments to Conjure are positional, for example the command name.
Another example of positional arguments is the path to a file required by the \sphinxcode{\sphinxupquote{conjure pretty}} command.
This argument can just be provided after the command name, like: \sphinxcode{\sphinxupquote{conjure pretty myfile.essence}}.

Non-positional arguments are provided using options.
Some options require an additional value to be provided.
Other options are flags and do not expect additional values.
For example the \sphinxcode{\sphinxupquote{conjure pretty}} command takes a flag called \sphinxcode{\sphinxupquote{-{-}remove-unused}}, which removes unused decision variables from the model before pretty printing it.
This option is a flag that takes no values.
However, the \sphinxcode{\sphinxupquote{conjure modelling}} command takes an option called \sphinxcode{\sphinxupquote{-{-}output-directory}}, which specifies the directory under which Conjure places its output files.
This option requires a value.

Options can have short or long names. Following the common convention, short option names are preceded by a single dash and long options names are preceded by two dashes.
For example \sphinxcode{\sphinxupquote{-{-}output-directory}} is a long name for an option, and \sphinxcode{\sphinxupquote{-o}} is a short name for the same option.

The general form of a Conjure run is as follows: \sphinxcode{\sphinxupquote{conjure {[}COMMAND{]} ... {[}OPTIONS{]}}}.

Following is the list of primary commands provided by Conjure.
They can be used to generate Essence’ models from Essence files, translate parameter files and solution files for a specific Essence’ model, and more.
\begin{description}
\item[{modelling}] \leavevmode
The main act. Given a problem specification in Essence, produce constraint programming models in Essence’.

\item[{translate-parameter}] \leavevmode
Refinement of parameter files written in Essence for a particular Essence’ model. The Essence’ model needs to be generated by Conjure.

\item[{translate-solution}] \leavevmode
Translation of solutions back to Essence.

\item[{validate-solution}] \leavevmode
Validating a solution.

\item[{solve}] \leavevmode
This is a combined mode, and it is available for convenience.
It runs conjure in the modelling mode followed by parameter refinement if required, then Savile Row + Minion to solve, and then solution translation.

\end{description}

If no primary command is provided, \sphinxcode{\sphinxupquote{modelling}} is assumed.

Conjure also supports a few additional commands on top of the primary commands listed above.
These commands are not required for the normal operation of the tool.
They are implemented to aid development and testing.
\begin{description}
\item[{pretty}] \leavevmode
Pretty print as Essence file to stdout. This mode can be used to view a binary Essence file in textual form.

\item[{diff}] \leavevmode
Diff on two Essence files. Works on models, parameters, and solutions.

\item[{type-check}] \leavevmode
Type-checking a single Essence file.

\item[{split}] \leavevmode
Split an Essence files to various smaller files. Useful for testing.

\item[{symmetry-detection}] \leavevmode
Dump some JSON to be used as input to ferret for symmetry detection.

\item[{parameter-generator}] \leavevmode
Generate an Essence model describing the instances of the problem class defined in the input Essence model. An error will be printed if the model has infinitely many instances.

\end{description}

Commands typically take additional arguments.
Each command provides a separate help message.
To see the command specific help message, run: \sphinxcode{\sphinxupquote{conjure COMMAND -{-}help}}.

\section{Help output}
\label{\detokenize{cli:help-output}}
The following is Conjure’s full help message for each command, provided for reference.
These messages may change between releases of Conjure.

\begin{sphinxVerbatim}[commandchars=\\\{\}]
Conjure: The Automated Constraint Modelling Tool

conjure [COMMAND] ... [OPTIONS]
  The command line interface of Conjure takes a command name as the first argument followed by more arguments depending
  on the command.
  This help text gives a list of the available commands.
  For details of a command, pass the \PYGZhy{}\PYGZhy{}help flag after the command name.
  For example: \PYGZsq{}conjure translate\PYGZhy{}solution \PYGZhy{}\PYGZhy{}help\PYGZsq{}

Common flags:
     \PYGZhy{}\PYGZhy{}help                                  Display help message
     \PYGZhy{}\PYGZhy{}version                               Print version information

conjure [modelling] [OPTIONS] ESSENCE\PYGZus{}FILE
  The main act. Given a problem specification in Essence, produce constraint programming models in Essence\PYGZsq{}.

Logging \PYGZam{} Output:
  \PYGZhy{}o \PYGZhy{}\PYGZhy{}output\PYGZhy{}directory=DIR                  Where to save generated models.
                                             Default value: \PYGZsq{}conjure\PYGZhy{}output\PYGZsq{}
     \PYGZhy{}\PYGZhy{}numbering\PYGZhy{}start=INT                   Starting value for output files.
                                             Default value: 1
     \PYGZhy{}\PYGZhy{}smart\PYGZhy{}filenames                       Use \PYGZdq{}smart names\PYGZdq{} for models.
                                             Directs Conjure to use the answers when producing a filename and to ignore
                                             the order of questions. Only useful if \PYGZsq{}f\PYGZsq{} is used for questions.
     \PYGZhy{}\PYGZhy{}log\PYGZhy{}level=LOGLEVEL                    Log level.
     \PYGZhy{}\PYGZhy{}verbose\PYGZhy{}trail                         Generate verbose trails.
     \PYGZhy{}\PYGZhy{}rewrites\PYGZhy{}trail                        Generate trails about the applied rewritings.
     \PYGZhy{}\PYGZhy{}log\PYGZhy{}rule\PYGZhy{}fails                        Generate logs for rule failures. (Caution: can be a lot!)
     \PYGZhy{}\PYGZhy{}log\PYGZhy{}rule\PYGZhy{}successes                    Generate logs for rule applications.
     \PYGZhy{}\PYGZhy{}log\PYGZhy{}rule\PYGZhy{}attempts                     Generate logs for rule attempts. (Caution: can be a lot!)
     \PYGZhy{}\PYGZhy{}log\PYGZhy{}choices                           Store the choices in a way that can be reused by \PYGZhy{}al
     \PYGZhy{}\PYGZhy{}output\PYGZhy{}format=FORMAT                  Format to use for output.
                                                 plain : default
                                                 binary: can be read by Conjure
                                                 json  : use to avoid parsing
     \PYGZhy{}\PYGZhy{}line\PYGZhy{}width=INT                        Line width for pretty printing.
                                             Default: 120
Model generation:
     \PYGZhy{}\PYGZhy{}responses=ITEM                        A comma separated list of integers.
                                             If provided, these will be used as the answers during interactive model
                                             generation instead of prompting the user.
     \PYGZhy{}\PYGZhy{}responses\PYGZhy{}representation=ITEM         A comma separated list of variable name : integer pairs.
                                             If provided, these will be used as the answers during interactive model
                                             generation instead of prompting the user for the variable representation
                                             questions.
                                             See \PYGZhy{}\PYGZhy{}dump\PYGZhy{}representations for a list of available representation options.
     \PYGZhy{}\PYGZhy{}estimate\PYGZhy{}number\PYGZhy{}of\PYGZhy{}models             Calculate (a lower bound on) the number of models, instead of running
                                             the usual modelling mode.
  \PYGZhy{}q \PYGZhy{}\PYGZhy{}strategy\PYGZhy{}q=STRATEGY                   Strategy for selecting the next question to answer. Options: f (for
                                             first), i (for interactive), r (for random), x (for all). Prepend a (for
                                             auto) to automatically skip when there is only one option at any point.
                                             Default value: f
  \PYGZhy{}a \PYGZhy{}\PYGZhy{}strategy\PYGZhy{}a=STRATEGY                   Strategy for selecting an answer. Same options as strategy\PYGZhy{}q.
                                              c picks the most \PYGZsq{}compact\PYGZsq{} option at every decision point.
                                              s picks the \PYGZsq{}sparsest\PYGZsq{} option at every decision point: useful for
                                             \PYGZhy{}\PYGZhy{}representations\PYGZhy{}givens
                                              l (for follow log) tries to pick given choices as far as possible
                                             Default value: ai
     \PYGZhy{}\PYGZhy{}representations=STRATEGY              Strategy for choosing a representation.
                                             Default value: same as \PYGZhy{}\PYGZhy{}strategy\PYGZhy{}a
     \PYGZhy{}\PYGZhy{}representations\PYGZhy{}finds=STRATEGY        Strategy for choosing a representation for a decision variable.
                                             Default value: same as \PYGZhy{}\PYGZhy{}representations
     \PYGZhy{}\PYGZhy{}representations\PYGZhy{}givens=STRATEGY       Strategy for choosing a representation for a parameter.
                                             Default value: s (for sparse)
     \PYGZhy{}\PYGZhy{}representations\PYGZhy{}auxiliaries=STRATEGY  Strategy for choosing a representation for an auxiliary variable.
                                             Default value: same as \PYGZhy{}\PYGZhy{}representations
     \PYGZhy{}\PYGZhy{}representations\PYGZhy{}quantifieds=STRATEGY  Strategy for choosing a representation for a quantified variable.
                                             Default value: same as \PYGZhy{}\PYGZhy{}representations
     \PYGZhy{}\PYGZhy{}representations\PYGZhy{}cuts=STRATEGY         Strategy for choosing a representation for cuts in \PYGZsq{}branching on\PYGZsq{}.
                                             Default value: same as \PYGZhy{}\PYGZhy{}representations
     \PYGZhy{}\PYGZhy{}channelling                           Whether to produce channelled models (true by default).
     \PYGZhy{}\PYGZhy{}representation\PYGZhy{}levels                 Whether to use built\PYGZhy{}in precedence levels when choosing representations.
                                             Used to cut down the number of generated models.
                                             Default: true
     \PYGZhy{}\PYGZhy{}seed=INT                              Random number generator seed.
     \PYGZhy{}\PYGZhy{}limit\PYGZhy{}models=INT                      Maximum number of models to generate.
     \PYGZhy{}\PYGZhy{}choices=FILE                          Choices to use for \PYGZhy{}al, either an eprime file (created by
                                             \PYGZhy{}\PYGZhy{}log\PYGZhy{}choices), or a json file.
General:
     \PYGZhy{}\PYGZhy{}limit\PYGZhy{}time=INT                        Limit in seconds of real time.

conjure translate\PYGZhy{}parameter [OPTIONS]
  Refinement of Essence parameter files for a particular Essence\PYGZsq{} model.
  The model needs to be generated by Conjure.

Flags:
     \PYGZhy{}\PYGZhy{}eprime=ESSENCE\PYGZus{}FILE                   An Essence\PYGZsq{} model generated by Conjure.
     \PYGZhy{}\PYGZhy{}essence\PYGZhy{}param=FILE                    An Essence parameter for the original problem specification.
     \PYGZhy{}\PYGZhy{}eprime\PYGZhy{}param=FILE                     An Essence\PYGZsq{} parameter matching the Essence\PYGZsq{} model.
                                             Default is \PYGZsq{}foo.eprime\PYGZhy{}param\PYGZsq{} if the Essence parameter file is named
                                             \PYGZsq{}foo.param\PYGZsq{}.
Logging \PYGZam{} Output:
     \PYGZhy{}\PYGZhy{}log\PYGZhy{}level=LOGLEVEL                    Log level.
     \PYGZhy{}\PYGZhy{}output\PYGZhy{}format=FORMAT                  Format to use for output.
                                                 plain : default
                                                 binary: can be read by Conjure
                                                 json  : use to avoid parsing
     \PYGZhy{}\PYGZhy{}line\PYGZhy{}width=INT                        Line width for pretty printing.
                                             Default: 120
General:
     \PYGZhy{}\PYGZhy{}limit\PYGZhy{}time=INT                        Limit in seconds of real time.

conjure translate\PYGZhy{}solution [OPTIONS]
  Translation of solutions back to Essence.

Flags:
     \PYGZhy{}\PYGZhy{}eprime=FILE                           An Essence\PYGZsq{} model generated by Conjure.
                                             Mandatory.
     \PYGZhy{}\PYGZhy{}essence\PYGZhy{}param=FILE                    An Essence parameter for the original problem specification.
                                             Mandatory.
     \PYGZhy{}\PYGZhy{}eprime\PYGZhy{}solution=FILE                  An Essence\PYGZsq{} solution for the corresponding Essence\PYGZsq{} model.
     \PYGZhy{}\PYGZhy{}essence\PYGZhy{}solution=FILE                 An Essence solution for the original problem specification.
                                             By default, its value is the value of \PYGZhy{}\PYGZhy{}eprime\PYGZhy{}solution with extensions
                                             replaced by \PYGZsq{}.solution\PYGZsq{}.
Logging \PYGZam{} Output:
     \PYGZhy{}\PYGZhy{}log\PYGZhy{}level=LOGLEVEL                    Log level.
     \PYGZhy{}\PYGZhy{}output\PYGZhy{}format=FORMAT                  Format to use for output.
                                                 plain : default
                                                 binary: can be read by Conjure
                                                 json  : use to avoid parsing
     \PYGZhy{}\PYGZhy{}line\PYGZhy{}width=INT                        Line width for pretty printing.
                                             Default: 120
General:
     \PYGZhy{}\PYGZhy{}limit\PYGZhy{}time=INT                        Limit in seconds of real time.

conjure validate\PYGZhy{}solution [OPTIONS]
  Validating a solution.

Flags:
     \PYGZhy{}\PYGZhy{}essence=ESSENCE\PYGZus{}FILE                  Problem specification in Essence.
     \PYGZhy{}\PYGZhy{}param=FILE                            Essence parameter file.
     \PYGZhy{}\PYGZhy{}solution=FILE                         Essence solution.
Logging \PYGZam{} Output:
     \PYGZhy{}\PYGZhy{}log\PYGZhy{}level=LOGLEVEL                    Log level.
     \PYGZhy{}\PYGZhy{}output\PYGZhy{}format=FORMAT                  Format to use for output.
                                                 plain : default
                                                 binary: can be read by Conjure
                                                 json  : use to avoid parsing
     \PYGZhy{}\PYGZhy{}line\PYGZhy{}width=INT                        Line width for pretty printing.
                                             Default: 120
General:
     \PYGZhy{}\PYGZhy{}limit\PYGZhy{}time=INT                        Limit in seconds of real time.

conjure solve [OPTIONS] ESSENCE\PYGZus{}FILE [PARAMETER\PYGZus{}FILE(s)]
  A combined mode for convenience.
  Runs Conjure in modelling mode followed by parameter translation if required, then Savile Row + Minion to solve, and
  then solution translation.

General:
     \PYGZhy{}\PYGZhy{}validate\PYGZhy{}solutions                    Enable solution validation.
     \PYGZhy{}\PYGZhy{}limit\PYGZhy{}time=INT                        Limit in seconds of real time.
     \PYGZhy{}\PYGZhy{}cgroups                               Setup and use cgroups when solving with Savile Row.
     \PYGZhy{}\PYGZhy{}number\PYGZhy{}of\PYGZhy{}solutions=ITEM              Number of solutions to find; \PYGZdq{}all\PYGZdq{} enumerates all solutions.
                                             Default: 1
     \PYGZhy{}\PYGZhy{}copy\PYGZhy{}solutions                        Whether to place a copy of solution(s) next to the Essence file or not.
                                             Default: on
Logging \PYGZam{} Output:
  \PYGZhy{}o \PYGZhy{}\PYGZhy{}output\PYGZhy{}directory=DIR                  Where to save generated models.
                                             Default value: \PYGZsq{}conjure\PYGZhy{}output\PYGZsq{}
     \PYGZhy{}\PYGZhy{}numbering\PYGZhy{}start=INT                   Starting value for output files.
                                             Default value: 1
     \PYGZhy{}\PYGZhy{}smart\PYGZhy{}filenames                       Use \PYGZdq{}smart names\PYGZdq{} for models.
                                             Directs Conjure to use the answers when producing a filename and to ignore
                                             the order of questions. Only useful if \PYGZsq{}f\PYGZsq{} is used for questions.
     \PYGZhy{}\PYGZhy{}solutions\PYGZhy{}in\PYGZhy{}one\PYGZhy{}file                 Place all solutions in a single file instead of generating a separate
                                             file per solution.
                                             Off by default.
     \PYGZhy{}\PYGZhy{}log\PYGZhy{}level=LOGLEVEL                    Log level.
     \PYGZhy{}\PYGZhy{}verbose\PYGZhy{}trail                         Generate verbose trails.
     \PYGZhy{}\PYGZhy{}rewrites\PYGZhy{}trail                        Generate trails about the applied rewritings.
     \PYGZhy{}\PYGZhy{}log\PYGZhy{}rule\PYGZhy{}fails                        Generate logs for rule failures. (Caution: can be a lot!)
     \PYGZhy{}\PYGZhy{}log\PYGZhy{}rule\PYGZhy{}successes                    Generate logs for rule applications.
     \PYGZhy{}\PYGZhy{}log\PYGZhy{}rule\PYGZhy{}attempts                     Generate logs for rule attempts. (Caution: can be a lot!)
     \PYGZhy{}\PYGZhy{}log\PYGZhy{}choices                           Store the choices in a way that can be reused by \PYGZhy{}al
     \PYGZhy{}\PYGZhy{}output\PYGZhy{}format=FORMAT                  Format to use for output.
                                                 plain : default
                                                 binary: can be read by Conjure
                                                 json  : use to avoid parsing
     \PYGZhy{}\PYGZhy{}line\PYGZhy{}width=INT                        Line width for pretty printing.
                                             Default: 120
Model generation:
     \PYGZhy{}\PYGZhy{}responses=ITEM                        A comma separated list of integers.
                                             If provided, these will be used as the answers during interactive model
                                             generation instead of prompting the user.
     \PYGZhy{}\PYGZhy{}responses\PYGZhy{}representation=ITEM         A comma separated list of variable name : integer pairs.
                                             If provided, these will be used as the answers during interactive model
                                             generation instead of prompting the user for the variable representation
                                             questions.
                                             See \PYGZhy{}\PYGZhy{}dump\PYGZhy{}representations for a list of available representation options.
  \PYGZhy{}q \PYGZhy{}\PYGZhy{}strategy\PYGZhy{}q=STRATEGY                   Strategy for selecting the next question to answer. Options: f (for
                                             first), i (for interactive), r (for random), x (for all). Prepend a (for
                                             auto) to automatically skip when there is only one option at any point.
                                             Default value: f
  \PYGZhy{}a \PYGZhy{}\PYGZhy{}strategy\PYGZhy{}a=STRATEGY                   Strategy for selecting an answer. Same options as strategy\PYGZhy{}q.
                                              c picks the most \PYGZsq{}compact\PYGZsq{} option at every decision point.
                                              s picks the \PYGZsq{}sparsest\PYGZsq{} option at every decision point: useful for
                                             \PYGZhy{}\PYGZhy{}representations\PYGZhy{}givens
                                              l (for follow log) tries to pick given choices as far as possible
                                             Default value: c
     \PYGZhy{}\PYGZhy{}representations=STRATEGY              Strategy for choosing a representation.
                                             Default value: same as \PYGZhy{}\PYGZhy{}strategy\PYGZhy{}a
     \PYGZhy{}\PYGZhy{}representations\PYGZhy{}finds=STRATEGY        Strategy for choosing a representation for a decision variable.
                                             Default value: same as \PYGZhy{}\PYGZhy{}representations
     \PYGZhy{}\PYGZhy{}representations\PYGZhy{}givens=STRATEGY       Strategy for choosing a representation for a parameter.
                                             Default value: s (for sparse)
     \PYGZhy{}\PYGZhy{}representations\PYGZhy{}auxiliaries=STRATEGY  Strategy for choosing a representation for an auxiliary variable.
                                             Default value: same as \PYGZhy{}\PYGZhy{}representations
     \PYGZhy{}\PYGZhy{}representations\PYGZhy{}quantifieds=STRATEGY  Strategy for choosing a representation for a quantified variable.
                                             Default value: same as \PYGZhy{}\PYGZhy{}representations
     \PYGZhy{}\PYGZhy{}representations\PYGZhy{}cuts=STRATEGY         Strategy for choosing a representation for cuts in \PYGZsq{}branching on\PYGZsq{}.
                                             Default value: same as \PYGZhy{}\PYGZhy{}representations
     \PYGZhy{}\PYGZhy{}channelling                           Whether to produce channelled models (true by default).
     \PYGZhy{}\PYGZhy{}representation\PYGZhy{}levels                 Whether to use built\PYGZhy{}in precedence levels when choosing representations.
                                             Used to cut down the number of generated models.
                                             Default: true
     \PYGZhy{}\PYGZhy{}seed=INT                              Random number generator seed.
     \PYGZhy{}\PYGZhy{}limit\PYGZhy{}models=INT                      Maximum number of models to generate.
     \PYGZhy{}\PYGZhy{}use\PYGZhy{}existing\PYGZhy{}models=FILE              File names of Essence\PYGZsq{} models generated beforehand.
                                             If given, Conjure skips the modelling phase and uses the existing models
                                             for solving.
                                             The models should be inside the output directory (See \PYGZhy{}o).
Options for other tools:
     \PYGZhy{}\PYGZhy{}savilerow\PYGZhy{}options=ITEM                Options passed to Savile Row.
     \PYGZhy{}\PYGZhy{}solver\PYGZhy{}options=ITEM                   Options passed to the backend solver.
     \PYGZhy{}\PYGZhy{}solver=ITEM                           Backend solver. Possible values:
                                              \PYGZhy{} minion (CP solver)
                                              \PYGZhy{} gecode (CP solver)
                                              \PYGZhy{} chuffed (CP solver)
                                              \PYGZhy{} glucose (SAT solver)
                                              \PYGZhy{} glucose\PYGZhy{}syrup (SAT solver)
                                              \PYGZhy{} lingeling (SAT solver)
                                              \PYGZhy{} minisat (SAT solver)
                                              \PYGZhy{} bc\PYGZus{}minisat\PYGZus{}all (AllSAT solver, only works with
                                             \PYGZhy{}\PYGZhy{}number\PYGZhy{}of\PYGZhy{}solutions=all)
                                              \PYGZhy{} nbc\PYGZus{}minisat\PYGZus{}all (AllSAT solver, only works with
                                             \PYGZhy{}\PYGZhy{}number\PYGZhy{}of\PYGZhy{}solutions=all)
                                              \PYGZhy{} open\PYGZhy{}wbo (MaxSAT solver, only works with optimisation problems)
                                             Default: minion

conjure ide [OPTIONS] ESSENCE\PYGZus{}FILE
  IDE support features for Conjure.
  Not intended for direct use.

Logging \PYGZam{} Output:
     \PYGZhy{}\PYGZhy{}log\PYGZhy{}level=LOGLEVEL                    Log level.
     \PYGZhy{}\PYGZhy{}line\PYGZhy{}width=INT                        Line width for pretty printing.
                                             Default: 120
General:
     \PYGZhy{}\PYGZhy{}limit\PYGZhy{}time=INT                        Limit in seconds of real time.
IDE Features:
     \PYGZhy{}\PYGZhy{}dump\PYGZhy{}declarations                     Print information about top level declarations.
     \PYGZhy{}\PYGZhy{}dump\PYGZhy{}representations                  List the available representations for decision variables and
                                             parameters.

conjure pretty [OPTIONS] ESSENCE\PYGZus{}FILE
  Pretty print as Essence file to stdout.
  This mode can be used to view a binary Essence file in textual form.

Flags:
     \PYGZhy{}\PYGZhy{}normalise\PYGZhy{}quantified                  Normalise the names of quantified variables.
     \PYGZhy{}\PYGZhy{}remove\PYGZhy{}unused                         Remove unused declarations.
Logging \PYGZam{} Output:
     \PYGZhy{}\PYGZhy{}log\PYGZhy{}level=LOGLEVEL                    Log level.
     \PYGZhy{}\PYGZhy{}output\PYGZhy{}format=FORMAT                  Format to use for output.
                                                 plain : default
                                                 binary: can be read by Conjure
                                                 json  : use to avoid parsing
     \PYGZhy{}\PYGZhy{}line\PYGZhy{}width=INT                        Line width for pretty printing.
                                             Default: 120
General:
     \PYGZhy{}\PYGZhy{}limit\PYGZhy{}time=INT                        Limit in seconds of real time.

conjure diff [OPTIONS] FILE FILE
  Diff on two Essence files. Works on models, parameters, and solutions.

Logging \PYGZam{} Output:
     \PYGZhy{}\PYGZhy{}log\PYGZhy{}level=LOGLEVEL                    Log level.
     \PYGZhy{}\PYGZhy{}output\PYGZhy{}format=FORMAT                  Format to use for output.
                                                 plain : default
                                                 binary: can be read by Conjure
                                                 json  : use to avoid parsing
     \PYGZhy{}\PYGZhy{}line\PYGZhy{}width=INT                        Line width for pretty printing.
                                             Default: 120
General:
     \PYGZhy{}\PYGZhy{}limit\PYGZhy{}time=INT                        Limit in seconds of real time.

conjure type\PYGZhy{}check [OPTIONS] ESSENCE\PYGZus{}FILE
  Type\PYGZhy{}checking a single Essence file.

Logging \PYGZam{} Output:
     \PYGZhy{}\PYGZhy{}log\PYGZhy{}level=LOGLEVEL                    Log level.
General:
     \PYGZhy{}\PYGZhy{}limit\PYGZhy{}time=INT                        Limit in seconds of real time.

conjure split [OPTIONS] ESSENCE\PYGZus{}FILE
  Split an Essence file to various smaller files. Useful for testing.

Logging \PYGZam{} Output:
  \PYGZhy{}o \PYGZhy{}\PYGZhy{}output\PYGZhy{}directory=DIR                  Where to save generated models.
                                             Default value: \PYGZsq{}conjure\PYGZhy{}output\PYGZsq{}
     \PYGZhy{}\PYGZhy{}log\PYGZhy{}level=LOGLEVEL                    Log level.
     \PYGZhy{}\PYGZhy{}output\PYGZhy{}format=FORMAT                  Format to use for output.
                                                 plain : default
                                                 binary: can be read by Conjure
                                                 json  : use to avoid parsing
     \PYGZhy{}\PYGZhy{}line\PYGZhy{}width=INT                        Line width for pretty printing.
                                             Default: 120
General:
     \PYGZhy{}\PYGZhy{}limit\PYGZhy{}time=INT                        Limit in seconds of real time.

conjure symmetry\PYGZhy{}detection [OPTIONS] ESSENCE\PYGZus{}FILE
  Dump some JSON to be used as input to ferret for symmetry detection.

Logging \PYGZam{} Output:
     \PYGZhy{}\PYGZhy{}json=JSON\PYGZus{}FILE                        Output JSON file.
                                             Default is \PYGZsq{}foo.essence\PYGZhy{}json\PYGZsq{}
                                             if the Essence file is named \PYGZsq{}foo.essence\PYGZsq{}
     \PYGZhy{}\PYGZhy{}log\PYGZhy{}level=LOGLEVEL                    Log level.
     \PYGZhy{}\PYGZhy{}output\PYGZhy{}format=FORMAT                  Format to use for output.
                                                 plain : default
                                                 binary: can be read by Conjure
                                                 json  : use to avoid parsing
     \PYGZhy{}\PYGZhy{}line\PYGZhy{}width=INT                        Line width for pretty printing.
                                             Default: 120
General:
     \PYGZhy{}\PYGZhy{}limit\PYGZhy{}time=INT                        Limit in seconds of real time.

conjure parameter\PYGZhy{}generator [OPTIONS] ESSENCE\PYGZus{}FILE
  Generate an Essence model describing the instances of the problem class defined in the input Essence model.
  An error will be printed if the model has infinitely many instances.

Logging \PYGZam{} Output:
     \PYGZhy{}\PYGZhy{}essence\PYGZhy{}out=FILE                      Output file path.
     \PYGZhy{}\PYGZhy{}log\PYGZhy{}level=LOGLEVEL                    Log level.
     \PYGZhy{}\PYGZhy{}output\PYGZhy{}format=FORMAT                  Format to use for output.
                                                 plain : default
                                                 binary: can be read by Conjure
                                                 json  : use to avoid parsing
     \PYGZhy{}\PYGZhy{}line\PYGZhy{}width=INT                        Line width for pretty printing.
                                             Default: 120
Integer bounds:
     \PYGZhy{}\PYGZhy{}MININT=INT                            The minimum integer value for the parameter values.
                                             Default: 0
     \PYGZhy{}\PYGZhy{}MAXINT=INT                            The maximum integer value for the parameter values.
                                             Default: 100
General:
     \PYGZhy{}\PYGZhy{}limit\PYGZhy{}time=INT                        Limit in seconds of real time.

conjure model\PYGZhy{}strengthening [OPTIONS] ESSENCE\PYGZus{}FILE
  Strengthen an Essence model as described in \PYGZdq{}Reformulating Essence Specifications for Robustness\PYGZdq{},
  which aims to make search faster.

Logging \PYGZam{} Output:
     \PYGZhy{}\PYGZhy{}essence\PYGZhy{}out=FILE                      Output file path.
     \PYGZhy{}\PYGZhy{}log\PYGZhy{}level=LOGLEVEL                    Log level.
     \PYGZhy{}\PYGZhy{}log\PYGZhy{}rule\PYGZhy{}successes                    Generate logs for rule applications.
     \PYGZhy{}\PYGZhy{}output\PYGZhy{}format=FORMAT                  Conjure\PYGZsq{}s output can be in multiple formats.
                                                 plain : The default
                                                 binary: A binary encoding of the Essence output.
                                                         It can be read back in by Conjure.
                                                 json  : A json encoding of the Essence output.
                                                         It can be used by other tools integrating with Conjure
                                                         in order to avoid having to parse textual Essence.
     \PYGZhy{}\PYGZhy{}line\PYGZhy{}width=INT                        Line width to use during pretty printing.
                                             Default: 120
General:
     \PYGZhy{}\PYGZhy{}limit\PYGZhy{}time=INT                        Time limit in seconds (real time).
\end{sphinxVerbatim}

\chapter{Features}
\label{\detokenize{features:features}}\label{\detokenize{features:id1}}\label{\detokenize{features::doc}}
This section lists some features of Conjure.

Some of these are due to features of Conjure’s input language Essence, and the need to support those. If you are not familiar with Essence, please see {\hyperref[\detokenize{essence:essence}]{\sphinxcrossref{\DUrole{std,std-ref}{Conjure’s input language: Essence}}}}.

\section{Problem classes}
\label{\detokenize{features:problem-classes}}
Often, when we think of problems we think of a \sphinxstyleemphasis{class} of problems rather than a single problem.
For example, Sudoku is a class of puzzles.
There are many different Sudoku \sphinxstyleemphasis{instances}, with different clues.
However, all Sudoku instances share the same set of rules.
Describing the puzzle of Sudoku to somebody who doesn’t know the rules of the game generally does not depend on a given set of clues.

Similarly, problem specifications in Essence are written for a class of problems instead of a single problem instance.
For Sudoku, the rules (everything on a row/column/sub-grid has to be distinct) are encoded once.
The clues are specified as \sphinxstyleemphasis{parameters} to the problem specification, together with appropriate assignment statements to incorporate the clues into the problem.

Many tools (solvers and/or modelling assistants) support this separation by having a parameterised problem specification in a file and separate data/parameter file specifying an instance of the problem.
Conjure uses \sphinxcode{\sphinxupquote{*.essence}} files for the problem specification, and \sphinxcode{\sphinxupquote{*.param}} files for the parameter file.

Although a lot of tools support this kind of a separation, they generally work by instantiating a problem specification before operating on it.
Conjure is different than most tools in this regard: it operates on parameterised problem specifications directly.
It reads in a parameterised Essence file, and outputs one or more parameterised Essence’ files.
To solve an Essence’ model provided by Conjure, Essence-level parameter files need to be translated to Essence’-level parameter files by running Conjure once per parameter file.

Savile Row accepts a parameterised model and a separate parameter file, and performs the instantiation.
The output model and the translated parameter file from Conjure can be directly used when running Savile Row.

\section{High level of abstraction}
\label{\detokenize{features:high-level-of-abstraction}}
Conjure’s input language is Essence.
Essence provides abstract domain types like sets, multi-sets, functions, sequences, relations, partitions, records, and variants.
These abstract domain types also support domain attributes like cardinality for set-like domains and injectivity/surjectivity for functions, to enable concise specification of a problem.
Essence also provides more primitive domain types like Booleans, integers, enumerated types, and matrices, that are supported by most CP solvers and modelling assistants.

In addition to abstract domain types, Essence also provides operators that operate on parameters or decision variables with abstract domains.
For example, set membership, subset, function inverse, and relation projection are provided to enable specification of problem constraints abstractly.

The high level of abstraction offered by Essence allows its users to specify problems without having to make a lot of low level \sphinxstyleemphasis{modelling decisions}.

\section{Arbitrarily nested types}
\label{\detokenize{features:arbitrarily-nested-types}}
The abstract domain types provided by Essence are domain constructors: they take another domain as an argument to construct a new domain.
For example the domain \sphinxcode{\sphinxupquote{set of D}} represents a set of values from the domain \sphinxcode{\sphinxupquote{D}}, and a \sphinxcode{\sphinxupquote{relation of (D1 * D2 * D3)}} represents a relation between values of domains \sphinxcode{\sphinxupquote{D1}}, \sphinxcode{\sphinxupquote{D2}}, and \sphinxcode{\sphinxupquote{D3}}.

Using these domain constructors, domains of arbitrary nesting can be created.
Conjure does not have a limit on the level of nesting in the domains.
But keep in mind: a several levels nested domain might look tiny whereas the combinatorial object it represents may be huge.

\section{Automatic symmetry breaking}
\label{\detokenize{features:automatic-symmetry-breaking}}
During its modelling process, a decision variable with an abstract domain type is \sphinxstyleemphasis{represented} using a collection of decision variables with more primitive domain types.
For example the domain \sphinxcode{\sphinxupquote{set (size n) of D}}, which represents a set of \sphinxcode{\sphinxupquote{n}} values from the domain \sphinxcode{\sphinxupquote{D}}, can be represented using the domain \sphinxcode{\sphinxupquote{matrix indexed by {[}int(1..n){]} of D}}.
Performing this modelling transformation requires rewriting the rest of the model.
Moreover, it introduces symmetry into the model, since a set implies a collection of distinct values whereas the matrix does not.
To break this symmetry, Conjure introduces strict ordering constraints on adjacent entries of the matrix.

This is one of the simplest examples of automated symmetry breaking performed by Conjure.
Conjure breaks all the symmetry introduced by modelling transformations like this one.

Another example is the domain \sphinxcode{\sphinxupquote{set of D}} without the explicit \sphinxcode{\sphinxupquote{size}} attribute.
Since the number of elements in this set is not known, Conjure cannot simply use a matrix to represent this domain.
There are multiple ways to represent this domain.
One representation is to use an integer to partition the entries of the matrix into two:
entries before the index pointed by this integer are regarded to be in the set, and
entries after this position are regarded to be irrelevant.

It is important to post constraints on the irrelevant entries to fix them to a certain value.
Not doing this introduces more symmetry.
Conjure breaks this kind of symmetry by introducing constraints to fix their values.

\section{Multiple models}
\label{\detokenize{features:multiple-models}}
Conjure is able to generate multiple Essence’ models starting from a single Essence problem specification.
Each model generated by Conjure can be used to solve the initial problem specified in Essence.

This feature is important because often a problem can be modelled in several ways, and it is difficult to know what a \sphinxstyleemphasis{good} model is for a given problem.
Constraint programming experts spend considerable amounts of time developing models.
It is common to create multiple models to compare how well they perform for a problem.
A good model is then chosen only after several models have been considered.

Moreover, a single good model may not even exist for certain classes of problems.
The choice of the model may depend on the instances we are interested in solving.

Lastly, instead of trying to pick a single good model a portfolio of models may be chosen with complementary strengths to exploit parallelism.

Conjure is able to produce multiple models mainly
\begin{itemize}
\item {} 
by having choices between multiple representations of decision variable domains, and

\item {} 
by having choices between translating constraint expressions in multiple ways.

\end{itemize}

Both domain representation and constraint translation mechanisms are implemented using a rule based system inside Conjure to ease the addition of new modelling idioms.

\section{Automated channelling}
\label{\detokenize{features:automated-channelling}}
While modelling a problem using constraint programming, it is often possible to model a certain decision using multiple encodings.
When different encodings with complementary strengths are available, experts can utilise this flexibility by using one encoding for parts of the formulation and another encoding for the rest of the formulation.
When multiple encodings of a single decision are used in a single model, \sphinxstyleemphasis{channelling} constraints are added to ensure consistency between encodings.

In Conjure, decision variables with abstract domain types can very often be represented in multiple ways.
For each occurrence of a decision variable, Conjure considers all representation options.
If a decision variable is used more than once, this means that the decision variable can be represented in multiple ways in a single Essence’ model.

When multiple representations are used, channelling constraints are generated by Conjure automatically.
These constraints make sure that different representations of the same abstract combinatorial object have the same abstract value.

\section{Extensibility}
\label{\detokenize{features:extensibility}}
The modelling transformations of Conjure are implemented using a rule-based system.

There are two main kinds of rules in Conjure:
\begin{description}
\item[{representations selection rules}] \leavevmode
to specify domain transformations,

\item[{expression refinement rules}] \leavevmode
to rewrite constraint expressions depending on their domain representations.

\end{description}

Moreover, Conjure contains a collection of \sphinxstylestrong{horizontal rules}, which are representation independent expression refinement rules.
Thanks to horizontal rules, the number of representation dependent expression refinement rules are kept to a small number.

Conjure’s architecture is designed to make adding both representation selection rules and expression refinement rules easy.

\section{Multiple target solvers}
\label{\detokenize{features:multiple-target-solvers}}
The ability to target multiple solvers is not a feature of Conjure by itself, but a benefit it gains thanks to being a part of a state-of-the-art constraint programming tool-chain.
Each Essence’ model generated by Conjure can be solved using \sphinxhref{http://savilerow.cs.st-andrews.ac.uk}{Savile Row} together with one of its target solvers.

Savile Row can directly target Minion, Gecode (via fzn-gecode), and any SAT solver that supports the DIMACS format.
It can also output Minizinc, and this output can be used to target a number of different solvers using the mzn2fzn tool.

\chapter{Conjure’s input language: Essence}
\label{\detokenize{essence:conjure-s-input-language-essence}}\label{\detokenize{essence:essence}}\label{\detokenize{essence::doc}}
Conjure works on problem specifications written in Essence.

This section gives a description of Essence.
A more thorough description can be found in the reference paper on Essence
\sphinxcite{zreferences:frisch2008essence}.

We adopt a BNF-style format to describe all the constructs of the language.
In the BNF format,
we use the “\#” character to denote comments,
we use double-quotes for terminal strings,
and we use a \sphinxcode{\sphinxupquote{list}} construct to indicate a list of syntax elements.

The \sphinxcode{\sphinxupquote{list}} construct has two variants:
\begin{enumerate}
\sphinxsetlistlabels{\arabic}{enumi}{enumii}{}{.}%
\item {} 
First variant takes two arguments where the first argument is the syntax of the items of the list and second argument is the item separator.

\item {} 
Second variant takes an additional third argument which indicates the surrounding bracket for the list. The third argument can be one of round brackets (\sphinxcode{\sphinxupquote{()}}), curly brackets (\sphinxcode{\sphinxupquote{\{\}}}), or square brackets (\sphinxcode{\sphinxupquote{{[}{]}}}).

\end{enumerate}

\begin{sphinxVerbatim}[commandchars=\\\{\}]
\PYG{k+kt}{ProblemSpecification} \PYG{o}{:=} \PYG{k}{list}\PYG{o}{(}\PYG{k+kt}{Statement}\PYG{o}{)}
\end{sphinxVerbatim}

A problem specification in Essence is composed of a list of statements.
Statements can declare decision variables, parameters or aliases.
They can also post constraints, conditions on parameter values and an objective statement.

The order of statements is largely insignificant, except in one case: names need to be declared before use.
For example a decision variable cannot be used before its declaration.

There are five kinds of statements in Essence.

\begin{sphinxVerbatim}[commandchars=\\\{\}]
\PYG{k+kt}{Statement} \PYG{o}{:=} \PYG{k+kt}{DeclarationStatement}
           \PYG{o}{\textbar{}} \PYG{k+kt}{BranchingStatement}
           \PYG{o}{\textbar{}} \PYG{k+kt}{SuchThatStatement}
           \PYG{o}{\textbar{}} \PYG{k+kt}{WhereStatement}
           \PYG{o}{\textbar{}} \PYG{k+kt}{ObjectiveStatement}
\end{sphinxVerbatim}

Every symbol must be declared before it is used, but otherwise statements can be listed in any order.
A problem specification can contain at most one branching statement.
A problem specification can contain at most one objective statement.

\section{Declarations}
\label{\detokenize{essence:declarations}}
\begin{sphinxVerbatim}[commandchars=\\\{\}]
\PYG{k+kt}{DeclarationStatement} \PYG{o}{:=} \PYG{k+kt}{FindStatement}
                      \PYG{o}{\textbar{}} \PYG{k+kt}{GivenStatement}
                      \PYG{o}{\textbar{}} \PYG{k+kt}{LettingStatement}
                      \PYG{o}{\textbar{}} \PYG{k+kt}{GivenEnum}
                      \PYG{o}{\textbar{}} \PYG{k+kt}{LettingEnum}
                      \PYG{o}{\textbar{}} \PYG{k+kt}{LettingUnnamed}
\end{sphinxVerbatim}

A declaration statement can be used to declare
a decision variable (\sphinxcode{\sphinxupquote{FindStatement}}),
a parameter (\sphinxcode{\sphinxupquote{GivenStatement}}),
an alias for an expression or a domain (\sphinxcode{\sphinxupquote{LettingStatement}}),
and enumerated or unnamed types.

\subsection{Declaring decision variables}
\label{\detokenize{essence:declaring-decision-variables}}
\begin{sphinxVerbatim}[commandchars=\\\{\}]
\PYG{k+kt}{FindStatement} \PYG{o}{:=} \PYG{l+s}{\PYGZdq{}}\PYG{l+s}{find}\PYG{l+s}{\PYGZdq{}} \PYG{k+kt}{Name} \PYG{l+s}{\PYGZdq{}}\PYG{l+s}{:}\PYG{l+s}{\PYGZdq{}} \PYG{k+kt}{Domain}
\end{sphinxVerbatim}

A decision variable is declared by using the keyword \sphinxcode{\sphinxupquote{find}}, followed by an identifier designating the name of the decision variable, followed by a colon symbol and the domain of the decision variable.
The domains of decision variables have to be finite.

This detail is omitted in the BNF above for simplicity, but a comma separated list of names may also be used to declare multiple decision variables with the same domain in a single find statement. This applies to all declaration statements.

\subsection{Declaring parameters}
\label{\detokenize{essence:declaring-parameters}}
\begin{sphinxVerbatim}[commandchars=\\\{\}]
\PYG{k+kt}{GivenStatement} \PYG{o}{:=} \PYG{l+s}{\PYGZdq{}}\PYG{l+s}{given}\PYG{l+s}{\PYGZdq{}} \PYG{k+kt}{Name} \PYG{l+s}{\PYGZdq{}}\PYG{l+s}{:}\PYG{l+s}{\PYGZdq{}} \PYG{k+kt}{Domain}
\end{sphinxVerbatim}

A parameter is declared in a similar way to decision variables. The only difference is the use of the keyword \sphinxcode{\sphinxupquote{given}} instead of the keyword \sphinxcode{\sphinxupquote{find}}.
Unlike decision variables, the domains of parameters do not have to be finite.

\subsection{Declaring aliases}
\label{\detokenize{essence:declaring-aliases}}
\begin{sphinxVerbatim}[commandchars=\\\{\}]
\PYG{k+kt}{LettingStatement} \PYG{o}{:=} \PYG{l+s}{\PYGZdq{}}\PYG{l+s}{letting}\PYG{l+s}{\PYGZdq{}} \PYG{k+kt}{Name} \PYG{l+s}{\PYGZdq{}}\PYG{l+s}{be}\PYG{l+s}{\PYGZdq{}} \PYG{k+kt}{Expression}
                  \PYG{o}{\textbar{}} \PYG{l+s}{\PYGZdq{}}\PYG{l+s}{letting}\PYG{l+s}{\PYGZdq{}} \PYG{k+kt}{Name} \PYG{l+s}{\PYGZdq{}}\PYG{l+s}{be}\PYG{l+s}{\PYGZdq{}} \PYG{l+s}{\PYGZdq{}}\PYG{l+s}{domain}\PYG{l+s}{\PYGZdq{}} \PYG{k+kt}{Domain}
\end{sphinxVerbatim}

An alias for an expression can be declared by using the keyword \sphinxcode{\sphinxupquote{letting}}, followed by the name of the alias, followed by the keyword \sphinxcode{\sphinxupquote{be}}, followed by an expression. Similarly, an alias for a domain can be declared by including the keyword \sphinxcode{\sphinxupquote{domain}} before writing the domain.

\begin{sphinxVerbatim}[commandchars=\\\{\}]
\PYG{k+kt}{letting} x \PYG{k+kt}{be} y + z
\PYG{k+kt}{letting} d \PYG{k+kt}{be} \PYG{k+kt}{domain} \PYG{k+kt}{set} \PYG{k+kt}{of} \PYG{k+kt}{int}\PYG{o}{(}a..b\PYG{o}{)}
\end{sphinxVerbatim}

In the example above \sphinxcode{\sphinxupquote{x}} is declared as an expression alias for \sphinxcode{\sphinxupquote{y + z}} and \sphinxcode{\sphinxupquote{d}} is declared as a domain alias for \sphinxcode{\sphinxupquote{set of int(a..b)}}.

\subsection{Declaring enumerated types}
\label{\detokenize{essence:declaring-enumerated-types}}
\begin{sphinxVerbatim}[commandchars=\\\{\}]
\PYG{k+kt}{GivenEnum} \PYG{o}{:=} \PYG{l+s}{\PYGZdq{}}\PYG{l+s}{given}\PYG{l+s}{\PYGZdq{}} \PYG{k+kt}{Name} \PYG{l+s}{\PYGZdq{}}\PYG{l+s}{new type enum}\PYG{l+s}{\PYGZdq{}}

\PYG{k+kt}{LettingEnum} \PYG{o}{:=} \PYG{l+s}{\PYGZdq{}}\PYG{l+s}{letting}\PYG{l+s}{\PYGZdq{}} \PYG{k+kt}{Name} \PYG{l+s}{\PYGZdq{}}\PYG{l+s}{be}\PYG{l+s}{\PYGZdq{}} \PYG{l+s}{\PYGZdq{}}\PYG{l+s}{new type enum}\PYG{l+s}{\PYGZdq{}} \PYG{k}{list}\PYG{o}{(}\PYG{k+kt}{Name}, \PYG{l+s}{\PYGZdq{}}\PYG{l+s}{,}\PYG{l+s}{\PYGZdq{}}, \PYG{l+s}{\PYGZdq{}}\PYG{l+s}{\PYGZob{}\PYGZcb{}}\PYG{l+s}{\PYGZdq{}}\PYG{o}{)}
\end{sphinxVerbatim}

Enumerated types can be declared in two ways: using a given-enum syntax or using a letting-enum syntax.

The given-enum syntax defers the specification of actual values of the enumerated type until instantiation.
With this syntax, an enumerated type can be declared by only giving its name in the problem specification file.
In a parameter file, values for the actual members of this type can be given.
This allows Conjure to produce a model independent of the values of the enumerated type and only substitute the actual values during parameter instantiation.

The letting-enum syntax can be used to declare an enumerated type directly in a problem specification as well.

\begin{sphinxVerbatim}[commandchars=\\\{\}]
\PYG{k+kt}{letting} direction \PYG{k+kt}{be} new type enum \PYG{o}{\PYGZob{}}North, East, South, West\PYG{o}{\PYGZcb{}}
\PYG{k+kt}{find} x,y : direction
such that x != y
\end{sphinxVerbatim}

In the example fragment above \sphinxcode{\sphinxupquote{direction}} is declared as an enumerated type with 4 members.
Two decision variables are declared using \sphinxcode{\sphinxupquote{direction}} as their domain and a constraint is posted on the values they can take.
Enumerated types support equality, ordering, and successor/predecessor operators; they do not support arithmetic operators.

When an enumerated type is declared, the elements of the type are listed in increasing order.

\subsection{Declaring unnamed types}
\label{\detokenize{essence:declaring-unnamed-types}}
\begin{sphinxVerbatim}[commandchars=\\\{\}]
\PYG{k+kt}{LettingUnnamed} \PYG{o}{:=} \PYG{l+s}{\PYGZdq{}}\PYG{l+s}{letting}\PYG{l+s}{\PYGZdq{}} \PYG{k+kt}{Name} \PYG{l+s}{\PYGZdq{}}\PYG{l+s}{be}\PYG{l+s}{\PYGZdq{}} \PYG{l+s}{\PYGZdq{}}\PYG{l+s}{new type of size}\PYG{l+s}{\PYGZdq{}} \PYG{k+kt}{Expression}
\end{sphinxVerbatim}

Unnamed types are a feature of Essence which allow succinct specification of certain types of symmetry.
An unnamed type is declared by giving it a name and a size (i.e. the number of elements in the type).
The members of an unnamed type cannot be referred to individually.
Typically constraints are posted using quantified variables over the whole domain.
Unnamed types only support equality operators; they do not support ordering or arithmetic operators.

\section{Branching statements}
\label{\detokenize{essence:branching-statements}}
\begin{sphinxVerbatim}[commandchars=\\\{\}]
\PYG{k+kt}{BranchingStatement} \PYG{o}{:=} \PYG{l+s}{\PYGZdq{}}\PYG{l+s}{branching}\PYG{l+s}{\PYGZdq{}} \PYG{l+s}{\PYGZdq{}}\PYG{l+s}{on}\PYG{l+s}{\PYGZdq{}} \PYG{k}{list}\PYG{o}{(}\PYG{k+kt}{BranchingOn}, \PYG{l+s}{\PYGZdq{}}\PYG{l+s}{,}\PYG{l+s}{\PYGZdq{}}, \PYG{l+s}{\PYGZdq{}}\PYG{l+s}{[]}\PYG{l+s}{\PYGZdq{}}\PYG{o}{)}

\PYG{k+kt}{BranchingOn} \PYG{o}{:=} \PYG{k+kt}{Name}
             \PYG{o}{\textbar{}} \PYG{k+kt}{Expression}
\end{sphinxVerbatim}

High level problem specification languages typically do not include lower level details such as directives specifying search order.
Essence is such a language, and the reference paper on Essence (\sphinxcite{zreferences:frisch2008essence}) does not include these search directives at all.

For pragmatic reasons Conjure supports search directives in the form of a branching-on statement, which takes a list of either variable names or expressions.
Decision variables in a branching-on statement are searched using a static value ordering.
Expressions can be used to introduce \sphinxstyleemphasis{cuts}; in which case when solving the model produced by Conjure, the solver is instructed to search for solutions satisfying the cut constraints first, and proceed to searching the rest of the search space later.

A problem specification can contain at most one branching statement.

\section{Constraints}
\label{\detokenize{essence:constraints}}
\begin{sphinxVerbatim}[commandchars=\\\{\}]
\PYG{k+kt}{SuchThatStatement} \PYG{o}{:=} \PYG{l+s}{\PYGZdq{}}\PYG{l+s}{such that}\PYG{l+s}{\PYGZdq{}} \PYG{k}{list}\PYG{o}{(}\PYG{k+kt}{Expression}, \PYG{l+s}{\PYGZdq{}}\PYG{l+s}{,}\PYG{l+s}{\PYGZdq{}}\PYG{o}{)}
\end{sphinxVerbatim}

Constraints are declared using the keyword sequence \sphinxcode{\sphinxupquote{such that}}, followed by a comma separated list of Boolean expressions.
The syntax for expressions is explained in section {\hyperref[\detokenize{essence:expressions}]{\sphinxcrossref{Expressions}}}.

\section{Instantiation conditions}
\label{\detokenize{essence:instantiation-conditions}}
\begin{sphinxVerbatim}[commandchars=\\\{\}]
\PYG{k+kt}{WhereStatement} \PYG{o}{:=} \PYG{l+s}{\PYGZdq{}}\PYG{l+s}{where}\PYG{l+s}{\PYGZdq{}} \PYG{k}{list}\PYG{o}{(}\PYG{k+kt}{Expression}, \PYG{l+s}{\PYGZdq{}}\PYG{l+s}{,}\PYG{l+s}{\PYGZdq{}}\PYG{o}{)}
\end{sphinxVerbatim}

Where statements are syntactically similar to {\hyperref[\detokenize{essence:constraints}]{\sphinxcrossref{constraints}}}, however they cannot refer to decision variables.
They can be used to post conditions on the parameters of the problem specification.
These conditions are checked during parameter instantiation.

\section{Objective statements}
\label{\detokenize{essence:objective-statements}}
\begin{sphinxVerbatim}[commandchars=\\\{\}]
\PYG{k+kt}{ObjectiveStatement} \PYG{o}{:=} \PYG{l+s}{\PYGZdq{}}\PYG{l+s}{minimising}\PYG{l+s}{\PYGZdq{}} \PYG{k+kt}{Expression}
                    \PYG{o}{\textbar{}} \PYG{l+s}{\PYGZdq{}}\PYG{l+s}{maximising}\PYG{l+s}{\PYGZdq{}} \PYG{k+kt}{Expression}
\end{sphinxVerbatim}

An objective can be declared by using either the keyword \sphinxcode{\sphinxupquote{minimising}} or the keyword \sphinxcode{\sphinxupquote{maximising}} followed by an integer expression.
A problem specification can have at most one objective statement.
If it has none it defines a satisfaction problem, if it has one it defines an optimisation problem.

A problem specification can contain at most one objective statement.

\section{Names}
\label{\detokenize{essence:names}}
The lexical rules for valid names in Essence are similar to those of most common languages.
A name consists of a sequence of non-whitespace alphanumeric characters (letters or digits) or underscores (\sphinxcode{\sphinxupquote{\_}}).
The first character of a valid name has to be a letter or an underscore.
Names are case-sensitive: Essence treats uppercase and lowercase versions of letters as distinct.

\section{Domains}
\label{\detokenize{essence:domains}}
\begin{sphinxVerbatim}[commandchars=\\\{\}]
\PYG{k+kt}{Domain} \PYG{o}{:=} \PYG{l+s}{\PYGZdq{}}\PYG{l+s}{bool}\PYG{l+s}{\PYGZdq{}}
        \PYG{o}{\textbar{}} \PYG{l+s}{\PYGZdq{}}\PYG{l+s}{int}\PYG{l+s}{\PYGZdq{}} \PYG{k}{list}\PYG{o}{(}Range, \PYG{l+s}{\PYGZdq{}}\PYG{l+s}{,}\PYG{l+s}{\PYGZdq{}}, \PYG{l+s}{\PYGZdq{}}\PYG{l+s}{()}\PYG{l+s}{\PYGZdq{}}\PYG{o}{)}
        \PYG{o}{\textbar{}} \PYG{l+s}{\PYGZdq{}}\PYG{l+s}{int}\PYG{l+s}{\PYGZdq{}} \PYG{l+s}{\PYGZdq{}}\PYG{l+s}{(}\PYG{l+s}{\PYGZdq{}} \PYG{k+kt}{Expression} \PYG{l+s}{\PYGZdq{}}\PYG{l+s}{)}\PYG{l+s}{\PYGZdq{}}
        \PYG{o}{\textbar{}} \PYG{k+kt}{Name} \PYG{k}{list}\PYG{o}{(}Range, \PYG{l+s}{\PYGZdq{}}\PYG{l+s}{,}\PYG{l+s}{\PYGZdq{}}, \PYG{l+s}{\PYGZdq{}}\PYG{l+s}{()}\PYG{l+s}{\PYGZdq{}}\PYG{o}{)} \PYG{c+cSingleline}{\PYGZsh{} the Name refers to an enumerated type}
        \PYG{o}{\textbar{}} \PYG{k+kt}{Name}                        \PYG{c+cSingleline}{\PYGZsh{} the Name refers to an unnamed type}
        \PYG{o}{\textbar{}} \PYG{l+s}{\PYGZdq{}}\PYG{l+s}{tuple}\PYG{l+s}{\PYGZdq{}} \PYG{k}{list}\PYG{o}{(}\PYG{k+kt}{Domain}, \PYG{l+s}{\PYGZdq{}}\PYG{l+s}{,}\PYG{l+s}{\PYGZdq{}}, \PYG{l+s}{\PYGZdq{}}\PYG{l+s}{()}\PYG{l+s}{\PYGZdq{}}\PYG{o}{)}
        \PYG{o}{\textbar{}} \PYG{l+s}{\PYGZdq{}}\PYG{l+s}{record}\PYG{l+s}{\PYGZdq{}} \PYG{k}{list}\PYG{o}{(}NameDomain, \PYG{l+s}{\PYGZdq{}}\PYG{l+s}{,}\PYG{l+s}{\PYGZdq{}}, \PYG{l+s}{\PYGZdq{}}\PYG{l+s}{\PYGZob{}\PYGZcb{}}\PYG{l+s}{\PYGZdq{}}\PYG{o}{)}
        \PYG{o}{\textbar{}} \PYG{l+s}{\PYGZdq{}}\PYG{l+s}{variant}\PYG{l+s}{\PYGZdq{}} \PYG{k}{list}\PYG{o}{(}NameDomain, \PYG{l+s}{\PYGZdq{}}\PYG{l+s}{,}\PYG{l+s}{\PYGZdq{}}, \PYG{l+s}{\PYGZdq{}}\PYG{l+s}{\PYGZob{}\PYGZcb{}}\PYG{l+s}{\PYGZdq{}}\PYG{o}{)}
        \PYG{o}{\textbar{}} \PYG{l+s}{\PYGZdq{}}\PYG{l+s}{matrix indexed by}\PYG{l+s}{\PYGZdq{}} \PYG{k}{list}\PYG{o}{(}\PYG{k+kt}{Domain}, \PYG{l+s}{\PYGZdq{}}\PYG{l+s}{,}\PYG{l+s}{\PYGZdq{}}, \PYG{l+s}{\PYGZdq{}}\PYG{l+s}{[]}\PYG{l+s}{\PYGZdq{}}\PYG{o}{)} \PYG{l+s}{\PYGZdq{}}\PYG{l+s}{of}\PYG{l+s}{\PYGZdq{}} \PYG{k+kt}{Domain}
        \PYG{o}{\textbar{}} \PYG{l+s}{\PYGZdq{}}\PYG{l+s}{set}\PYG{l+s}{\PYGZdq{}} \PYG{k}{list}\PYG{o}{(}Attribute, \PYG{l+s}{\PYGZdq{}}\PYG{l+s}{,}\PYG{l+s}{\PYGZdq{}}, \PYG{l+s}{\PYGZdq{}}\PYG{l+s}{()}\PYG{l+s}{\PYGZdq{}}\PYG{o}{)} \PYG{l+s}{\PYGZdq{}}\PYG{l+s}{of}\PYG{l+s}{\PYGZdq{}} \PYG{k+kt}{Domain}
        \PYG{o}{\textbar{}} \PYG{l+s}{\PYGZdq{}}\PYG{l+s}{mset}\PYG{l+s}{\PYGZdq{}} \PYG{k}{list}\PYG{o}{(}Attribute, \PYG{l+s}{\PYGZdq{}}\PYG{l+s}{,}\PYG{l+s}{\PYGZdq{}}, \PYG{l+s}{\PYGZdq{}}\PYG{l+s}{()}\PYG{l+s}{\PYGZdq{}}\PYG{o}{)} \PYG{l+s}{\PYGZdq{}}\PYG{l+s}{of}\PYG{l+s}{\PYGZdq{}} \PYG{k+kt}{Domain}
        \PYG{o}{\textbar{}} \PYG{l+s}{\PYGZdq{}}\PYG{l+s}{function}\PYG{l+s}{\PYGZdq{}} \PYG{k}{list}\PYG{o}{(}Attribute, \PYG{l+s}{\PYGZdq{}}\PYG{l+s}{,}\PYG{l+s}{\PYGZdq{}}, \PYG{l+s}{\PYGZdq{}}\PYG{l+s}{()}\PYG{l+s}{\PYGZdq{}}\PYG{o}{)} \PYG{k+kt}{Domain} \PYG{l+s}{\PYGZdq{}}\PYG{l+s}{\PYGZhy{}\PYGZhy{}\PYGZgt{}}\PYG{l+s}{\PYGZdq{}} \PYG{k+kt}{Domain}
        \PYG{o}{\textbar{}} \PYG{l+s}{\PYGZdq{}}\PYG{l+s}{sequence}\PYG{l+s}{\PYGZdq{}} \PYG{k}{list}\PYG{o}{(}Attribute, \PYG{l+s}{\PYGZdq{}}\PYG{l+s}{,}\PYG{l+s}{\PYGZdq{}}, \PYG{l+s}{\PYGZdq{}}\PYG{l+s}{()}\PYG{l+s}{\PYGZdq{}}\PYG{o}{)} \PYG{l+s}{\PYGZdq{}}\PYG{l+s}{of}\PYG{l+s}{\PYGZdq{}} \PYG{k+kt}{Domain}
        \PYG{o}{\textbar{}} \PYG{l+s}{\PYGZdq{}}\PYG{l+s}{relation}\PYG{l+s}{\PYGZdq{}} \PYG{k}{list}\PYG{o}{(}Attribute, \PYG{l+s}{\PYGZdq{}}\PYG{l+s}{,}\PYG{l+s}{\PYGZdq{}}, \PYG{l+s}{\PYGZdq{}}\PYG{l+s}{()}\PYG{l+s}{\PYGZdq{}}\PYG{o}{)} \PYG{l+s}{\PYGZdq{}}\PYG{l+s}{of}\PYG{l+s}{\PYGZdq{}} \PYG{k}{list}\PYG{o}{(}\PYG{k+kt}{Domain}, \PYG{l+s}{\PYGZdq{}}\PYG{l+s}{*}\PYG{l+s}{\PYGZdq{}}, \PYG{l+s}{\PYGZdq{}}\PYG{l+s}{()}\PYG{l+s}{\PYGZdq{}}\PYG{o}{)}
        \PYG{o}{\textbar{}} \PYG{l+s}{\PYGZdq{}}\PYG{l+s}{partition}\PYG{l+s}{\PYGZdq{}} \PYG{k}{list}\PYG{o}{(}Attribute, \PYG{l+s}{\PYGZdq{}}\PYG{l+s}{,}\PYG{l+s}{\PYGZdq{}}, \PYG{l+s}{\PYGZdq{}}\PYG{l+s}{()}\PYG{l+s}{\PYGZdq{}}\PYG{o}{)} \PYG{l+s}{\PYGZdq{}}\PYG{l+s}{from}\PYG{l+s}{\PYGZdq{}} \PYG{k+kt}{Domain}

Range \PYG{o}{:=} \PYG{k+kt}{Expression}
       \PYG{o}{\textbar{}} \PYG{k+kt}{Expression} \PYG{l+s}{\PYGZdq{}}\PYG{l+s}{..}\PYG{l+s}{\PYGZdq{}}
       \PYG{o}{\textbar{}} \PYG{l+s}{\PYGZdq{}}\PYG{l+s}{..}\PYG{l+s}{\PYGZdq{}} \PYG{k+kt}{Expression}
       \PYG{o}{\textbar{}} \PYG{k+kt}{Expression} \PYG{l+s}{\PYGZdq{}}\PYG{l+s}{..}\PYG{l+s}{\PYGZdq{}} \PYG{k+kt}{Expression}

Attribute \PYG{o}{:=} \PYG{k+kt}{Name}
           \PYG{o}{\textbar{}} \PYG{k+kt}{Name} \PYG{k+kt}{Expression}

NameDomain \PYG{o}{:=} \PYG{k+kt}{Name} \PYG{l+s}{\PYGZdq{}}\PYG{l+s}{:}\PYG{l+s}{\PYGZdq{}} \PYG{k+kt}{Domain}
\end{sphinxVerbatim}

Essence contains a rich selection of domain constructors, which can be used in an arbitrarily nested fashion to create domains for problem parameters, decision variables, quantified expressions and comprehensions.
Quantified expressions and comprehensions are explained under {\hyperref[\detokenize{essence:expressions}]{\sphinxcrossref{Expressions}}}.

Domains can be finite or infinite, but infinite domains can only be used when declaring of problem parameters.
The domains for both decision variables and quantified variables have to be finite.

Some kinds of domains can take an optional list of attributes.
An attribute is either a label or a label with an associated value.
Different kinds of domains take different attributes.

Multiple attributes can be used in a single domain.
Using contradicting values for the attribute values may result in an empty domain.

In the following, each kind of domain is described in a subsection of its own.

\subsection{Boolean domains}
\label{\detokenize{essence:boolean-domains}}
The Boolean domain is denoted with the keyword \sphinxcode{\sphinxupquote{bool}} and has two values: \sphinxcode{\sphinxupquote{false}} and \sphinxcode{\sphinxupquote{true}}.
The Boolean domain is ordered with \sphinxcode{\sphinxupquote{false}} preceding \sphinxcode{\sphinxupquote{true}}.
It is not currently possible to specify an objective with respect to a Boolean value.
If \sphinxcode{\sphinxupquote{a}} is a Boolean variable to minimise or maximise in the objective, use \sphinxcode{\sphinxupquote{toInt(a)}} instead (see {\hyperref[\detokenize{essence:type-conversion-operators}]{\sphinxcrossref{Type conversion operators}}}).

\subsection{Integer domains}
\label{\detokenize{essence:integer-domains}}
An integer domain is denoted by the keyword \sphinxcode{\sphinxupquote{int}}, followed by a list of integer ranges inside round brackets.
The list of ranges is optional, if omitted the integer domain denotes the infinite domain of all integers.

An integer range is either a single integer, or a list of sequential integers with a given lower and upper bound.
The bounds can be omitted to create an open range, but note that using open ranges inside an integer domain declaration creates an infinite domain.

Integer domains can also be constructed using a single set expression inside the round brackets, instead of a list of ranges.
The integer domain contains all members of the set in this case.
Note that the set expression cannot contain references to decision variables if this syntax is used.

Values in an integer domain should be in the range -2**62+1 to 2**62-1 as values outside this range may trigger errors in Savile Row or Minion, and lead to Conjure unexpectedly but silently deducing unsatisfiability.
Intermediate values in an integer expression must also be inside this range.

\subsection{Enumerated domains}
\label{\detokenize{essence:enumerated-domains}}
Enumerated types are declared using the syntax given in {\hyperref[\detokenize{essence:declaring-enumerated-types}]{\sphinxcrossref{Declaring enumerated types}}}.

An enumerated domain is denoted by using the name of the enumerated type, followed by a list of ranges inside round brackets.
The list of ranges is optional, if omitted the enumerated domain denotes the finite domain containing all values of the enumerated type.

A range is either a single value (member of the enumerated type), or a list of sequential values with a given lower and upper bound.
The bounds can be omitted to create an open range, when an open range is used the omitted bound is considered to be the same as the corresponding bound of the enumerated type.

\subsection{Unnamed domains}
\label{\detokenize{essence:unnamed-domains}}
Unnamed types are declared using the syntax given in {\hyperref[\detokenize{essence:declaring-unnamed-types}]{\sphinxcrossref{Declaring unnamed types}}}.

An unnamed domain is denoted by using the name of the unnamed type.
It does not take a list of ranges to limit the values in the domain, an unnamed domain always contains all values in the corresponding unnamed type.

\subsection{Tuple domains}
\label{\detokenize{essence:tuple-domains}}
Tuple is a domain constructor, it takes a list of domains as arguments.
Tuples can be of arbitrary arity.

A tuple domain is denoted by the keyword \sphinxcode{\sphinxupquote{tuple}}, followed by a list of domains separated by commas inside round brackets.
The keyword \sphinxcode{\sphinxupquote{tuple}} is optional for tuples of arity greater or equal to 2.

When needed, domains inside a tuple are referred to using their positions.
In an n-arity tuple, the position of the first domain is 1, and the position of the last domain is n.

To explicitly specify a tuple, use a list of values inside round brackets, preceded by the keyword \sphinxcode{\sphinxupquote{tuple}}.

\begin{sphinxVerbatim}[commandchars=\\\{\}]
\PYG{k+kt}{letting} s \PYG{k+kt}{be} tuple\PYG{o}{(}\PYG{o}{)}
\PYG{k+kt}{letting} t \PYG{k+kt}{be} tuple\PYG{o}{(}0,1,1,1\PYG{o}{)}
\end{sphinxVerbatim}

\subsection{Record domains}
\label{\detokenize{essence:record-domains}}
Record is a domain constructor, it takes a list of name-domain pairs as arguments.
Records can be of arbitrary arity.

A record domain is denoted by the keyword \sphinxcode{\sphinxupquote{record}}, followed by a list of name-domain pairs separated by commas inside curly brackets.

Records are very similar to tuples; except they use labels for their components instead of positions.
When needed, domains inside a record are referred to using their labels.

\subsection{Variant domains}
\label{\detokenize{essence:variant-domains}}
Variant is a domain constructor, it takes a list of name-domain pairs as arguments.
Variants can be of arbitrary arity.

A variant domain is denoted by the keyword \sphinxcode{\sphinxupquote{variant}}, followed by a list of name-domain pairs separated by commas inside curly brackets.

Variants are similar to records but with a very important distinction.
A member of a record domain contains a value for each component of the record, however
a member of a variant domain contains a value for only one of the components of the variant.

Variant domains are similar to \sphinxhref{http://en.wikipedia.org/wiki/Tagged\_union}{tagged unions} in other programming languages.

\subsection{Matrix domains}
\label{\detokenize{essence:matrix-domains}}
Matrix is a domain constructor, it takes a list of domains for its indices and a domain for the entries of the matrix.
Matrices can be of arbitrary dimensionality (greater than 0).

A matrix domain is denoted by the keywords \sphinxcode{\sphinxupquote{matrix indexed by}},
followed by a list of domains separated by commas inside square brackets,
followed by the keyword \sphinxcode{\sphinxupquote{of}}, and another domain.

A matrix can be indexed only by integer, Boolean, or enumerated domains.

Matrix domains are the most basic container-like domains in Essence.
They are used when the decision variable or the problem parameter does not have any further relevant structure.
Using another kind of domain is more appropriate for most problem specifications in Essence.

Matrix domains are not ordered, but matrices can be compared using the equality operators.

To explicitly specify a matrix, use a list of values inside square brackets.

\begin{sphinxVerbatim}[commandchars=\\\{\}]
\PYG{k+kt}{letting} M \PYG{k+kt}{be} [0,1,0,\PYGZhy{}1]
\PYG{k+kt}{letting} N \PYG{k+kt}{be} [[0,1],[0,\PYGZhy{}1]]
\end{sphinxVerbatim}

\subsection{Set domains}
\label{\detokenize{essence:set-domains}}
Set is a domain constructor, it takes a domain as argument denoting the domain of the members of the set.

A set domain is denoted by the keyword \sphinxcode{\sphinxupquote{set}},
followed by an optional comma separated list of set attributes,
followed by the keyword \sphinxcode{\sphinxupquote{of}}, and the domain for members of the set.

Set attributes are all related to cardinality: \sphinxcode{\sphinxupquote{size}}, \sphinxcode{\sphinxupquote{minSize}}, and \sphinxcode{\sphinxupquote{maxSize}}.

To explicitly specify a set, use a list of values inside curly brackets.
Values only appear once in the set; if repeated values are specified then they are ignored.

\begin{sphinxVerbatim}[commandchars=\\\{\}]
\PYG{k+kt}{letting} S \PYG{k+kt}{be} \PYG{o}{\PYGZob{}}1,0,1\PYG{o}{\PYGZcb{}}
\end{sphinxVerbatim}

\subsection{Multi-set domains}
\label{\detokenize{essence:multi-set-domains}}
Multi-set is a domain constructor, it takes a domain as argument denoting the domain of the members of the multi-set.

A multi-set domain is denoted by the keyword \sphinxcode{\sphinxupquote{mset}},
followed by an optional comma separated list of multi-set attributes,
followed by the keyword \sphinxcode{\sphinxupquote{of}}, and the domain for members of the multi-set.

There are two groups of multi-set attributes:
\begin{enumerate}
\sphinxsetlistlabels{\arabic}{enumi}{enumii}{}{.}%
\item {} 
Related to cardinality: \sphinxcode{\sphinxupquote{size}}, \sphinxcode{\sphinxupquote{minSize}}, and \sphinxcode{\sphinxupquote{maxSize}}.

\item {} 
Related to number of occurrences of values in the multi-set: \sphinxcode{\sphinxupquote{minOccur}}, and \sphinxcode{\sphinxupquote{maxOccur}}.

\end{enumerate}

Since a multi-set domain is infinite without a \sphinxcode{\sphinxupquote{size}}, \sphinxcode{\sphinxupquote{maxSize}}, or \sphinxcode{\sphinxupquote{maxOccur}} attribute, one of these attributes is mandatory to define a finite domain.

To explicitly specify a multi-set, use a list of values inside round brackets, preceded by the keyword \sphinxcode{\sphinxupquote{mset}}.
Values may appear multiple times in a multi-set.

\begin{sphinxVerbatim}[commandchars=\\\{\}]
\PYG{k+kt}{letting} S \PYG{k+kt}{be} mset\PYG{o}{(}0,1,1,1\PYG{o}{)}
\end{sphinxVerbatim}

\subsection{Function domains}
\label{\detokenize{essence:function-domains}}
Function is a domain constructor, it takes two domains as arguments denoting the \sphinxstyleemphasis{defined} and the \sphinxstyleemphasis{range} sets of the function.
It is important to take note that we are using \sphinxstyleemphasis{defined} to mean the domain of the function, and \sphinxstyleemphasis{range} to mean the codomain.

A function domain is denoted by the keyword \sphinxcode{\sphinxupquote{function}},
followed by an optional comma separated list of function attributes,
followed by the two domains separated by an arrow symbol: \sphinxcode{\sphinxupquote{-{-}\textgreater{}}}.

There are three groups of function attributes:
\begin{enumerate}
\sphinxsetlistlabels{\arabic}{enumi}{enumii}{}{.}%
\item {} 
Related to cardinality: \sphinxcode{\sphinxupquote{size}}, \sphinxcode{\sphinxupquote{minSize}}, and \sphinxcode{\sphinxupquote{maxSize}}.

\item {} 
Related to function properties: \sphinxcode{\sphinxupquote{injective}}, \sphinxcode{\sphinxupquote{surjective}}, and \sphinxcode{\sphinxupquote{bijective}}.

\item {} 
Related to partiality: \sphinxcode{\sphinxupquote{total}}.

\end{enumerate}

Cardinality attributes take arguments, but the rest of the arguments do not.
Function domains are partial by default, and using the \sphinxcode{\sphinxupquote{total}} attribute makes them total.

To explicitly specify a function, use a list of assignments, each of the form \sphinxcode{\sphinxupquote{input -{-}\textgreater{} value}}, inside round brackets and preceded by the keyword \sphinxcode{\sphinxupquote{function}}.

\begin{sphinxVerbatim}[commandchars=\\\{\}]
\PYG{k+kt}{letting} f \PYG{k+kt}{be} function\PYG{o}{(}0\PYGZhy{}\PYGZhy{}\PYGZgt{}1,1\PYGZhy{}\PYGZhy{}\PYGZgt{}0\PYG{o}{)}
\end{sphinxVerbatim}

\subsection{Sequence domains}
\label{\detokenize{essence:sequence-domains}}
Sequence is a domain constructor, it takes a domain as argument denoting the domain of the members of the sequence.

A sequence is denoted by the keyword \sphinxcode{\sphinxupquote{sequence}},
followed by an optional comma separated list of sequence attributes,
followed by the keyword \sphinxcode{\sphinxupquote{of}}, and the domain for members of the sequence.

There are 2 groups of sequence attributes:
\begin{enumerate}
\sphinxsetlistlabels{\arabic}{enumi}{enumii}{}{.}%
\item {} 
Related to cardinality: \sphinxcode{\sphinxupquote{size}}, \sphinxcode{\sphinxupquote{minSize}}, and \sphinxcode{\sphinxupquote{maxSize}}.

\item {} 
Related to function-like properties: \sphinxcode{\sphinxupquote{injective}}, \sphinxcode{\sphinxupquote{surjective}}, and \sphinxcode{\sphinxupquote{bijective}}.

\end{enumerate}

Cardinality attributes take arguments, but the rest of the arguments do not.
Sequence domains are total by default, hence they do not take a separate \sphinxcode{\sphinxupquote{total}} attribute.

Sequences are indexed by a contiguous list of increasing integers, beginning at 1.

To explicitly specify a sequence, use a list of values inside round brackets, preceded by the keyword \sphinxcode{\sphinxupquote{sequence}}.

\begin{sphinxVerbatim}[commandchars=\\\{\}]
\PYG{k+kt}{letting} s \PYG{k+kt}{be} sequence\PYG{o}{(}1,0,\PYGZhy{}1,2\PYG{o}{)}
\PYG{k+kt}{letting} t \PYG{k+kt}{be} sequence\PYG{o}{(}\PYG{o}{)} \PYGZdl{} empty sequence
\end{sphinxVerbatim}

\subsection{Relation domains}
\label{\detokenize{essence:relation-domains}}
Relation is a domain constructor, it takes a list of domains as arguments.
Relations can be of arbitrary arity.

A relation domain is denoted by the keyword \sphinxcode{\sphinxupquote{relation}},
followed by an optional comma separated list of relation attributes,
followed by the keyword \sphinxcode{\sphinxupquote{of}}, and a list of domains separated by the \sphinxcode{\sphinxupquote{*}} symbol inside round brackets.

There are 2 groups of relation attributes:
\begin{enumerate}
\sphinxsetlistlabels{\arabic}{enumi}{enumii}{}{.}%
\item {} 
Related to cardinality: \sphinxcode{\sphinxupquote{size}}, \sphinxcode{\sphinxupquote{minSize}}, and \sphinxcode{\sphinxupquote{maxSize}}.

\item {} 
Binary relation attributes: \sphinxcode{\sphinxupquote{reflexive}}, \sphinxcode{\sphinxupquote{irreflexive}}, \sphinxcode{\sphinxupquote{coreflexive}}, \sphinxcode{\sphinxupquote{symmetric}}, \sphinxcode{\sphinxupquote{antiSymmetric}}, \sphinxcode{\sphinxupquote{aSymmetric}}, \sphinxcode{\sphinxupquote{transitive}}, \sphinxcode{\sphinxupquote{total}}, \sphinxcode{\sphinxupquote{connex}}, \sphinxcode{\sphinxupquote{Euclidean}}, \sphinxcode{\sphinxupquote{serial}}, \sphinxcode{\sphinxupquote{equivalence}}, \sphinxcode{\sphinxupquote{partialOrder}}.

\end{enumerate}

The binary relation attributes are only applicable to relations of arity 2, and are between two identical domains.

To explicitly specify a relation, use a list of tuples, enclosed by round brackets and preceded by the keyword \sphinxcode{\sphinxupquote{relation}}.
All the tuples must be of the same type.

\begin{sphinxVerbatim}[commandchars=\\\{\}]
\PYG{k+kt}{letting} R \PYG{k+kt}{be} relation\PYG{o}{(}\PYG{o}{(}1,1,0\PYG{o}{)},\PYG{o}{(}1,0,1\PYG{o}{)},\PYG{o}{(}0,1,1\PYG{o}{)}\PYG{o}{)}
\end{sphinxVerbatim}

\subsection{Partition domains}
\label{\detokenize{essence:partition-domains}}
Partition is a domain constructor, it takes a domain as an argument denoting the members in the partition.

A partition is denoted by the keyword \sphinxcode{\sphinxupquote{partition}},
followed by an optional comma separated list of partition attributes,
followed by the keyword \sphinxcode{\sphinxupquote{from}}, and the domain for the members in the partition.

There are 3 groups of partition attributes:
\begin{enumerate}
\sphinxsetlistlabels{\arabic}{enumi}{enumii}{}{.}%
\item {} 
Related to the number of parts in the partition: \sphinxcode{\sphinxupquote{numParts}}, \sphinxcode{\sphinxupquote{minNumParts}}, and \sphinxcode{\sphinxupquote{maxNumParts}}.

\item {} 
Related to the cardinality of each part in the partition: \sphinxcode{\sphinxupquote{partSize}}, \sphinxcode{\sphinxupquote{minPartSize}}, and \sphinxcode{\sphinxupquote{maxPartSize}}.

\item {} 
Partition properties: \sphinxcode{\sphinxupquote{regular}}.

\end{enumerate}

The first and second groups of attributes are related to number of parts and cardinalities of each part in the partition.
The \sphinxcode{\sphinxupquote{regular}} attribute forces each part to be of the same cardinality without specifying the actual number of parts or cardinalities of each part.

\section{Types}
\label{\detokenize{essence:types}}
Essence is a statically typed language.
A declaration \textendash{} whether it is a decision variable, a problem parameter or a quantified variable \textendash{} has an associated domain.
From its domain, a type can be calculated.

A type is obtained from a domain by
removing attributes (from set, multi-set, function, sequence, relation, and partition domains),
and removing bounds (from integer and enumerated domains).

In the expression language of Essence, each operator has a typing rule associated with it.
These typing rules are used to both type check expression fragments and to calculate the types of resulting expressions.

For example, the arithmetic operator \sphinxcode{\sphinxupquote{+}} requires two arguments both of which are integers, and the resulting expression is also an integer.
So if \sphinxcode{\sphinxupquote{a}}, and \sphinxcode{\sphinxupquote{b}} are integers \sphinxcode{\sphinxupquote{a + b}} is also an integer.
Conjure gives a type error otherwise.

Using these typing rules every Essence expression can be checked for type correctness statically.

\section{Expressions}
\label{\detokenize{essence:expressions}}
\begin{sphinxVerbatim}[commandchars=\\\{\}]
\PYG{k+kt}{Expression} \PYG{o}{:=} \PYG{k+kt}{Literal}
            \PYG{o}{\textbar{}} \PYG{k+kt}{Name}
            \PYG{o}{\textbar{}} \PYG{k+kt}{Quantification}
            \PYG{o}{\textbar{}} \PYG{k+kt}{Comprehension} \PYG{k+kt}{Expression} [GeneratorOrCondition]
            \PYG{o}{\textbar{}} \PYG{k+kt}{Operator}

\PYG{k+kt}{Operator} \PYG{o}{:=} ...
\end{sphinxVerbatim}

(In preparation)

\subsection{Matrix indexing}
\label{\detokenize{essence:matrix-indexing}}
A list is a one-dimensional matrix indexed by an integer, starting at 1.
Matrices of dimension k are implemented by a list of matrices of dimension k-1.

\begin{sphinxVerbatim}[commandchars=\\\{\}]
\PYG{k+kt}{letting} D1 \PYG{k+kt}{be} \PYG{k+kt}{domain} matrix indexed by [int\PYG{o}{(}1..2\PYG{o}{)},int\PYG{o}{(}1..5\PYG{o}{)}] \PYG{k+kt}{of} \PYG{k+kt}{int}\PYG{o}{(}\PYGZhy{}1..1\PYG{o}{)}
\PYG{k+kt}{letting} E \PYG{k+kt}{be} \PYG{k+kt}{domain} matrix indexed by [int\PYG{o}{(}1..5\PYG{o}{)}] \PYG{k+kt}{of} \PYG{k+kt}{int}\PYG{o}{(}\PYGZhy{}1..1\PYG{o}{)}
\PYG{k+kt}{letting} D2 \PYG{k+kt}{be} \PYG{k+kt}{domain} matrix indexed by [int\PYG{o}{(}1..2\PYG{o}{)}] \PYG{k+kt}{of} E
\PYG{k+kt}{find} A : D1 such that A[1] = [\PYGZhy{}1,1,1,0,1], A[2] = [1,1,1,1,1]
\PYG{k+kt}{find} B : D2 such that B[1] = A[1], B[2] = [0,0,0,0,0]
\PYG{k+kt}{letting} C \PYG{k+kt}{be} [[\PYGZhy{}1,1,1,0,1],[0,0,0,0,0]]
\PYG{k+kt}{letting} a \PYG{k+kt}{be} A[1][1] = \PYGZhy{}1                   \PYGZdl{} true
\PYG{k+kt}{letting} b \PYG{k+kt}{be} A[1,1] = \PYGZhy{}1                    \PYGZdl{} true
\PYG{k+kt}{letting} c \PYG{k+kt}{be} C[1] = [\PYGZhy{}1,1,1,0,1]            \PYGZdl{} true
\PYG{k+kt}{letting} d \PYG{k+kt}{be} B[1] = C[1]                    \PYGZdl{} true
\PYG{k+kt}{letting} e \PYG{k+kt}{be} [A[1],B[2]] = C                \PYGZdl{} true
\PYG{k+kt}{letting} f \PYG{k+kt}{be} B = C                          \PYGZdl{} true
\PYG{k+kt}{letting} F \PYG{k+kt}{be} \PYG{k+kt}{domain} matrix indexed by [int\PYG{o}{(}1..6\PYG{o}{)}] \PYG{k+kt}{of} bool
\PYG{k+kt}{find} g : F such that g = [a,b,c,d,e,f] \PYGZdl{} [true,true,true,true,true,true]
\end{sphinxVerbatim}

\subsection{Tuple indexing}
\label{\detokenize{essence:tuple-indexing}}
Tuples are indexed by a constant integer, starting at 1.
Attempting to access a tuple element via an index that is negative, zero, or too large for the tuple, results in an error.

\begin{sphinxVerbatim}[commandchars=\\\{\}]
\PYG{k+kt}{letting} s \PYG{k+kt}{be} tuple\PYG{o}{(}0,1,1,0\PYG{o}{)}
\PYG{k+kt}{letting} t \PYG{k+kt}{be} tuple\PYG{o}{(}0,0,0,1\PYG{o}{)}
\PYG{k+kt}{find} a : bool such that a = \PYG{o}{(}s[1] = t[1]\PYG{o}{)} \PYGZdl{} true
\end{sphinxVerbatim}

\subsection{Arithmetic operators}
\label{\detokenize{essence:arithmetic-operators}}
Essence supports the four usual arithmetic operators
\begin{quote}

\begin{DUlineblock}{0em}
\item[] \sphinxcode{\sphinxupquote{+}}  \sphinxcode{\sphinxupquote{-}}  \sphinxcode{\sphinxupquote{*}}  \sphinxcode{\sphinxupquote{/}}
\end{DUlineblock}
\end{quote}

and also the modulo operator \sphinxcode{\sphinxupquote{\%}}, exponentiation \sphinxcode{\sphinxupquote{**}}.
These all take two arguments and are expressed in infix notation.

There is also the unary prefix operator \sphinxcode{\sphinxupquote{-}} for negation, the unary postfix operator \sphinxcode{\sphinxupquote{!}} for the factorial function, and the absolute value operator \sphinxcode{\sphinxupquote{\textbar{}x\textbar{}}}.

The arithmetic operators have the usual precedence: the factorial operator is applied first, then exponentiation, then negation, then the multiplication, division, and modulo operators, and finally addition and subtraction.

Exponentiation associates to the right, other binary operators to the left.

\subsubsection{Division}
\label{\detokenize{essence:division}}
Division returns an integer, and the following relationship holds when \sphinxcode{\sphinxupquote{x}} and \sphinxcode{\sphinxupquote{y}} are integers and \sphinxcode{\sphinxupquote{y}} is not zero:
\begin{quote}

\begin{DUlineblock}{0em}
\item[] \sphinxcode{\sphinxupquote{(x \% y) + y*(x / y) = x}}
\end{DUlineblock}
\end{quote}

whenever \sphinxcode{\sphinxupquote{y}} is not zero.
\sphinxcode{\sphinxupquote{x / 0}} and \sphinxcode{\sphinxupquote{x \% 0}} are expressions that do not have a defined value.
Division by zero may lead to unsatisfiability but is not flagged by either Conjure or Savile Row as an error.

\subsubsection{Factorial}
\label{\detokenize{essence:factorial}}
Both \sphinxcode{\sphinxupquote{factorial(x)}} and \sphinxcode{\sphinxupquote{x!}} denote the product of all positive integers up to \sphinxcode{\sphinxupquote{x}}, with \sphinxcode{\sphinxupquote{x! = 1}} whenever \sphinxcode{\sphinxupquote{x \textless{}= 0}}.
The factorial operator cannot be used directly in expressions involving decision variables, so the following

\begin{sphinxVerbatim}[commandchars=\\\{\}]
\PYG{k+kt}{find} z : \PYG{k+kt}{int}\PYG{o}{(}\PYGZhy{}1..13\PYG{o}{)}
such that \PYG{o}{(}z! \PYGZgt{} 2**28\PYG{o}{)}
\end{sphinxVerbatim}

is flagged as an error.
However, the following does work:

\begin{sphinxVerbatim}[commandchars=\\\{\}]
\PYG{k+kt}{find} z : \PYG{k+kt}{int}\PYG{o}{(}\PYGZhy{}1..13\PYG{o}{)}
such that \PYG{o}{(}exists x : \PYG{k+kt}{int}\PYG{o}{(}\PYGZhy{}1..13\PYG{o}{)} . \PYG{o}{(}x! \PYGZgt{} 2**28\PYG{o}{)} /\PYGZbs{} \PYG{o}{(}z=x\PYG{o}{)}\PYG{o}{)}
\end{sphinxVerbatim}

\subsubsection{Powers}
\label{\detokenize{essence:powers}}
When \sphinxcode{\sphinxupquote{x}} is an integer and \sphinxcode{\sphinxupquote{y}} is a positive integer, then \sphinxcode{\sphinxupquote{x**y}} denotes \sphinxcode{\sphinxupquote{x}} raised to the \sphinxcode{\sphinxupquote{y}}-th power.
When \sphinxcode{\sphinxupquote{y}} is a negative integer, \sphinxcode{\sphinxupquote{x**y}} is flagged by Savile Row as an error (this includes \sphinxcode{\sphinxupquote{1**(-1)}}).
Conjure does not flag negative powers as errors.
The relationship
\begin{quote}

\begin{DUlineblock}{0em}
\item[] \sphinxcode{\sphinxupquote{x ** y = x*(x**(y-1))}}
\end{DUlineblock}
\end{quote}

holds for all integers \sphinxcode{\sphinxupquote{x}} and positive integers \sphinxcode{\sphinxupquote{y}}.
This means that \sphinxcode{\sphinxupquote{x**0}} is always 1, whatever the value of \sphinxcode{\sphinxupquote{x}}.

\subsubsection{Negation}
\label{\detokenize{essence:negation}}
The unary operator \sphinxcode{\sphinxupquote{-}} denotes negation; when \sphinxcode{\sphinxupquote{x}} is an integer then \sphinxcode{\sphinxupquote{-{-}x = x}} is always true.

\subsubsection{Absolute value}
\label{\detokenize{essence:absolute-value}}
When \sphinxcode{\sphinxupquote{x}} is an integer, \sphinxcode{\sphinxupquote{\textbar{}x\textbar{}}} denotes the absolute value of \sphinxcode{\sphinxupquote{x}}.
The relationship
\begin{quote}

\begin{DUlineblock}{0em}
\item[] \sphinxcode{\sphinxupquote{(2*toInt(x \textgreater{}= 0) - 1)*x = \textbar{}x\textbar{}}}
\end{DUlineblock}
\end{quote}

holds for all integers \sphinxcode{\sphinxupquote{x}} such that \sphinxcode{\sphinxupquote{\textbar{}x\textbar{} \textless{}= 2**62-2}}.
Integers outside this range may be flagged as an error by Savile Row and/or Minion.

\subsection{Comparisons}
\label{\detokenize{essence:comparisons}}
The inline binary comparison operators \sphinxcode{\sphinxupquote{=}}  \sphinxcode{\sphinxupquote{!=}}  \sphinxcode{\sphinxupquote{\textless{}}}  \sphinxcode{\sphinxupquote{\textless{}=}}  \sphinxcode{\sphinxupquote{\textgreater{}}}  \sphinxcode{\sphinxupquote{\textless{}=}} can be used to compare two expressions.

The equality operators \sphinxcode{\sphinxupquote{=}} and \sphinxcode{\sphinxupquote{!=}} can be applied to compare two expressions, both taking values in the same domain.
Equality operators are supported for all types.

The equality operators have the same precedence as other logical operators.
This may lead to unintended unsatisfiability or introducing inadvertent solutions.
This is illustrated in the following example, where there are two possible solutions.

\begin{sphinxVerbatim}[commandchars=\\\{\}]
\PYG{k+kt}{find} a : bool such that a = false  \PYGZbs{}/ true  \PYGZdl{} true or false
\PYG{k+kt}{find} b : bool such that b = \PYG{o}{(}false \PYGZbs{}/ true\PYG{o}{)} \PYGZdl{} true
\end{sphinxVerbatim}

The inline binary comparison operators \sphinxcode{\sphinxupquote{\textless{}}}  \sphinxcode{\sphinxupquote{\textless{}=}}  \sphinxcode{\sphinxupquote{\textgreater{}}}  \sphinxcode{\sphinxupquote{\textless{}=}} can be used to compare expressions taking values in an ordered domain.
The expressions must both be integer, both Boolean or both enumerated types.

\begin{sphinxVerbatim}[commandchars=\\\{\}]
\PYG{k+kt}{letting} direction \PYG{k+kt}{be} new type enum \PYG{o}{\PYGZob{}}North, East, South, West\PYG{o}{\PYGZcb{}}
\PYG{k+kt}{find} a : bool such that a = \PYG{o}{(}\PYG{o}{(}North \PYGZlt{} South\PYG{o}{)} /\PYGZbs{} \PYG{o}{(}South \PYGZlt{} West\PYG{o}{)}\PYG{o}{)}  \PYGZdl{} true
\PYG{k+kt}{find} b : bool such that b = \PYG{o}{(}false \PYGZlt{}= true\PYG{o}{)} \PYGZdl{} true
\end{sphinxVerbatim}

The inline binary comparison operators
\begin{quote}

\begin{DUlineblock}{0em}
\item[] \sphinxcode{\sphinxupquote{\textless{}lex}} \sphinxcode{\sphinxupquote{\textless{}=lex}} \sphinxcode{\sphinxupquote{\textgreater{}lex}} \sphinxcode{\sphinxupquote{\textgreater{}=lex}}
\end{DUlineblock}
\end{quote}

test whether their arguments have the specified relative lexicographic order.

\subsection{Logical operators}
\label{\detokenize{essence:logical-operators}}

\begin{savenotes}\sphinxattablestart
\centering
\begin{tabulary}{\linewidth}[t]{|T|T|}
\hline

\sphinxcode{\sphinxupquote{/\textbackslash{}}}
&
and
\\
\hline
\sphinxcode{\sphinxupquote{\textbackslash{}/}}
&
or
\\
\hline
\sphinxcode{\sphinxupquote{-\textgreater{}}}
&
implication
\\
\hline
\sphinxcode{\sphinxupquote{\textless{}-\textgreater{}}}
&
if and only if
\\
\hline
\sphinxcode{\sphinxupquote{!}}
&
negation
\\
\hline
\end{tabulary}
\par
\sphinxattableend\end{savenotes}

Logical operators operate on Boolean valued expressions, returning a Boolean value \sphinxcode{\sphinxupquote{false}} or \sphinxcode{\sphinxupquote{true}}.
Negation is unary prefix, the others are binary inline.
The \sphinxcode{\sphinxupquote{and}}, \sphinxcode{\sphinxupquote{or}} and \sphinxcode{\sphinxupquote{xor}} operators can be applied to sets or lists of Boolean values (see {\hyperref[\detokenize{essence:list-combining-operators}]{\sphinxcrossref{List combining operators}}} for details).
Note that \sphinxcode{\sphinxupquote{\textless{}-}} is not a logical operator, but is used in list comprehension syntax.

\subsection{Set operators}
\label{\detokenize{essence:set-operators}}
The following set operators return Boolean values indicating whether a specific relationship holds:

\begin{savenotes}\sphinxattablestart
\centering
\begin{tabulary}{\linewidth}[t]{|T|T|}
\hline

\sphinxcode{\sphinxupquote{in}}
&
test if element is in set
\\
\hline
\sphinxcode{\sphinxupquote{subset}}
&
test if first set is strictly contained in second set
\\
\hline
\sphinxcode{\sphinxupquote{subsetEq}}
&
test if first set is contained in second set
\\
\hline
\sphinxcode{\sphinxupquote{supset}}
&
test if first set strictly contains second set
\\
\hline
\sphinxcode{\sphinxupquote{supsetEq}}
&
test if first set contains second set
\\
\hline
\end{tabulary}
\par
\sphinxattableend\end{savenotes}

These binary inline operators operate on sets and return a set:

\begin{savenotes}\sphinxattablestart
\centering
\begin{tabulary}{\linewidth}[t]{|T|T|}
\hline

\sphinxcode{\sphinxupquote{intersect}}
&
set of elements in both sets
\\
\hline
\sphinxcode{\sphinxupquote{union}}
&
set of elements in either of the sets
\\
\hline
\end{tabulary}
\par
\sphinxattableend\end{savenotes}

The following unary operator operates on a set and returns a set:

\begin{savenotes}\sphinxattablestart
\centering
\begin{tabulary}{\linewidth}[t]{|T|T|}
\hline

\sphinxcode{\sphinxupquote{powerSet}}
&
set of all subsets of a set (including the empty set)
\\
\hline
\end{tabulary}
\par
\sphinxattableend\end{savenotes}

When \sphinxcode{\sphinxupquote{S}} is a set, then \sphinxcode{\sphinxupquote{\textbar{}S\textbar{}}} denotes the non-negative integer that is the cardinality of \sphinxcode{\sphinxupquote{S}} (the number of elements in \sphinxcode{\sphinxupquote{S}}).
When \sphinxcode{\sphinxupquote{S}} and \sphinxcode{\sphinxupquote{T}} are sets, \sphinxcode{\sphinxupquote{S - T}} denotes their set difference, the set of elements of \sphinxcode{\sphinxupquote{S}} that do not occur in \sphinxcode{\sphinxupquote{T}}.

Examples:

\begin{sphinxVerbatim}[commandchars=\\\{\}]
\PYG{k+kt}{find} a : bool such that a = \PYG{o}{(}1 in \PYG{o}{\PYGZob{}}0,1\PYG{o}{\PYGZcb{}}\PYG{o}{)} \PYGZdl{} true
\PYG{k+kt}{find} b : bool such that b = \PYG{o}{(}\PYG{o}{\PYGZob{}}0,1\PYG{o}{\PYGZcb{}} subset \PYG{o}{\PYGZob{}}0,1\PYG{o}{\PYGZcb{}}\PYG{o}{)} \PYGZdl{} false
\PYG{k+kt}{find} c : bool such that c = \PYG{o}{(}\PYG{o}{\PYGZob{}}0,1\PYG{o}{\PYGZcb{}} subsetEq \PYG{o}{\PYGZob{}}0,1\PYG{o}{\PYGZcb{}}\PYG{o}{)} \PYGZdl{} true
\PYG{k+kt}{find} d : bool such that d = \PYG{o}{(}\PYG{o}{\PYGZob{}}0,1\PYG{o}{\PYGZcb{}} supset \PYG{o}{\PYGZob{}}\PYG{o}{\PYGZcb{}}\PYG{o}{)} \PYGZdl{} true
\PYG{k+kt}{find} e : bool such that e = \PYG{o}{(}\PYG{o}{\PYGZob{}}0,1\PYG{o}{\PYGZcb{}} supsetEq \PYG{o}{\PYGZob{}}1,0\PYG{o}{\PYGZcb{}}\PYG{o}{)} \PYGZdl{} true
\PYG{k+kt}{find} A : \PYG{k+kt}{set} \PYG{k+kt}{of} \PYG{k+kt}{int}\PYG{o}{(}0..6\PYG{o}{)} such that A = \PYG{o}{\PYGZob{}}1,2,3\PYG{o}{\PYGZcb{}} intersect \PYG{o}{\PYGZob{}}3,4\PYG{o}{\PYGZcb{}} \PYGZdl{} \PYG{o}{\PYGZob{}}3\PYG{o}{\PYGZcb{}}
\PYG{k+kt}{find} B : \PYG{k+kt}{set} \PYG{k+kt}{of} \PYG{k+kt}{int}\PYG{o}{(}0..6\PYG{o}{)} such that B = \PYG{o}{\PYGZob{}}1,2,3\PYG{o}{\PYGZcb{}} union \PYG{o}{\PYGZob{}}3,4\PYG{o}{\PYGZcb{}} \PYGZdl{} \PYG{o}{\PYGZob{}}1,2,3,4\PYG{o}{\PYGZcb{}}
\PYG{k+kt}{find} S : \PYG{k+kt}{set} \PYG{k+kt}{of} \PYG{k+kt}{set} \PYG{k+kt}{of} \PYG{k+kt}{int}\PYG{o}{(}0..2\PYG{o}{)} such that S = powerSet\PYG{o}{(}\PYG{o}{\PYGZob{}}0\PYG{o}{\PYGZcb{}}\PYG{o}{)} \PYGZdl{} \PYG{o}{\PYGZob{}}\PYG{o}{\PYGZob{}}\PYG{o}{\PYGZcb{}},\PYG{o}{\PYGZob{}}0\PYG{o}{\PYGZcb{}}\PYG{o}{\PYGZcb{}}
\PYG{k+kt}{find} x : \PYG{k+kt}{int}\PYG{o}{(}0..9\PYG{o}{)} such that x = \PYG{o}{\textbar{}}\PYG{o}{\PYGZob{}}0,1,2,1,2,1\PYG{o}{\PYGZcb{}}\PYG{o}{\textbar{}} \PYGZdl{} 3
\PYG{k+kt}{find} T : \PYG{k+kt}{set} \PYG{k+kt}{of} \PYG{k+kt}{int}\PYG{o}{(}0..9\PYG{o}{)} such that T = \PYG{o}{\PYGZob{}}0,1,2\PYG{o}{\PYGZcb{}} \PYGZhy{} \PYG{o}{\PYGZob{}}2,3\PYG{o}{\PYGZcb{}} \PYGZdl{} \PYG{o}{\PYGZob{}}0,1\PYG{o}{\PYGZcb{}}
\end{sphinxVerbatim}

\subsection{Sequence operators}
\label{\detokenize{essence:sequence-operators}}
For two sequences \sphinxcode{\sphinxupquote{s}} and \sphinxcode{\sphinxupquote{t}}, \sphinxcode{\sphinxupquote{s subsequence t}} tests whether the list of values taken by \sphinxcode{\sphinxupquote{s}} occurs in the same order in the list of values taken by \sphinxcode{\sphinxupquote{t}}, and \sphinxcode{\sphinxupquote{s substring t}} tests whether the list of values taken by \sphinxcode{\sphinxupquote{s}} occurs in the same order and contiguously in the list of values taken by \sphinxcode{\sphinxupquote{t}}.

When \sphinxcode{\sphinxupquote{S}} is a sequence, then \sphinxcode{\sphinxupquote{\textbar{}S\textbar{}}} denotes the number of elements in \sphinxcode{\sphinxupquote{S}}.

\begin{sphinxVerbatim}[commandchars=\\\{\}]
\PYG{k+kt}{letting} s \PYG{k+kt}{be} sequence\PYG{o}{(}1,1\PYG{o}{)}
\PYG{k+kt}{letting} t \PYG{k+kt}{be} sequence\PYG{o}{(}2,1,3,1\PYG{o}{)}
\PYG{k+kt}{find} a : bool such that s subsequence t \PYGZdl{} true
\PYG{k+kt}{find} b : bool such that s substring t \PYGZdl{} false
\PYG{k+kt}{find} c : \PYG{k+kt}{int}\PYG{o}{(}1..10\PYG{o}{)} such that c = \PYG{o}{\textbar{}}t\textbar{} \PYGZdl{} 4
\end{sphinxVerbatim}

\subsection{Enumerated type operators}
\label{\detokenize{essence:enumerated-type-operators}}

\begin{savenotes}\sphinxattablestart
\centering
\begin{tabulary}{\linewidth}[t]{|T|T|}
\hline

\sphinxcode{\sphinxupquote{pred}}
&
predecessor of this element in an enumerated type
\\
\hline
\sphinxcode{\sphinxupquote{succ}}
&
successor of this element in an enumerated type
\\
\hline
\end{tabulary}
\par
\sphinxattableend\end{savenotes}

Enumerated types are ordered, so they support comparisons and the operators \sphinxtitleref{max} and \sphinxtitleref{min}.

\begin{sphinxVerbatim}[commandchars=\\\{\}]
\PYG{k+kt}{letting} D \PYG{k+kt}{be} new type enum \PYG{o}{\PYGZob{}} North, East, South, West \PYG{o}{\PYGZcb{}}
\PYG{k+kt}{find} a : D such that a = succ\PYG{o}{(}East\PYG{o}{)} \PYGZdl{} South
\PYG{k+kt}{find} b : bool such that b = \PYG{o}{(}max\PYG{o}{(}[North, South]\PYG{o}{)} \PYGZgt{} East\PYG{o}{)} \PYGZdl{} true
\end{sphinxVerbatim}

\subsection{Multiset operators}
\label{\detokenize{essence:multiset-operators}}
The following operators take a single argument:

\begin{savenotes}\sphinxattablestart
\centering
\begin{tabulary}{\linewidth}[t]{|T|T|}
\hline

\sphinxcode{\sphinxupquote{hist}}
&
histogram of multi-set/matrix
\\
\hline
\sphinxcode{\sphinxupquote{max}}
&
largest element in ordered set/multi-set/domain/list
\\
\hline
\sphinxcode{\sphinxupquote{min}}
&
smallest element in ordered set/multi-set/domain/list
\\
\hline
\end{tabulary}
\par
\sphinxattableend\end{savenotes}

The following operator takes two arguments:

\begin{savenotes}\sphinxattablestart
\centering
\begin{tabulary}{\linewidth}[t]{|T|T|}
\hline

\sphinxcode{\sphinxupquote{freq}}
&
counts occurrences of element in multi-set/matrix
\\
\hline
\end{tabulary}
\par
\sphinxattableend\end{savenotes}

Examples:

\begin{sphinxVerbatim}[commandchars=\\\{\}]
\PYG{k+kt}{letting} S \PYG{k+kt}{be} mset\PYG{o}{(}0,1,\PYGZhy{}1,1\PYG{o}{)}
\PYG{k+kt}{find} x : \PYG{k+kt}{int}\PYG{o}{(}0..1\PYG{o}{)} such that freq\PYG{o}{(}S,x\PYG{o}{)} = 2 \PYGZdl{} 1
\PYG{k+kt}{find} y : \PYG{k+kt}{int}\PYG{o}{(}\PYGZhy{}2..2\PYG{o}{)} such that y = max\PYG{o}{(}S\PYG{o}{)} \PYGZhy{} min\PYG{o}{(}S\PYG{o}{)} \PYGZdl{} 2
\PYG{k+kt}{find} z : \PYG{k+kt}{int}\PYG{o}{(}\PYGZhy{}2..2\PYG{o}{)} such that z = max\PYG{o}{(}[1,2]\PYG{o}{)} \PYGZdl{} 2
\end{sphinxVerbatim}

\subsection{Type conversion operators}
\label{\detokenize{essence:type-conversion-operators}}

\begin{savenotes}\sphinxattablestart
\centering
\begin{tabulary}{\linewidth}[t]{|T|T|}
\hline

\sphinxcode{\sphinxupquote{toInt}}
&
maps \sphinxcode{\sphinxupquote{true}} to 1, \sphinxcode{\sphinxupquote{false}} to 0
\\
\hline
\sphinxcode{\sphinxupquote{toMSet}}
&
set/relation/function to multi-set
\\
\hline
\sphinxcode{\sphinxupquote{toRelation}}
&
function to relation; \sphinxcode{\sphinxupquote{function(a -{-}\textgreater{} b)}} becomes
\sphinxcode{\sphinxupquote{relation((a,b))}}
\\
\hline
\sphinxcode{\sphinxupquote{toSet}}
&
multi-set/relation/function to set; \sphinxcode{\sphinxupquote{mset(0,0,1)}}
becomes \sphinxcode{\sphinxupquote{\{0,1\}}}
\\
\hline
\end{tabulary}
\par
\sphinxattableend\end{savenotes}

It is currently not possible to use an operator to directly invert \sphinxcode{\sphinxupquote{toRelation}} or \sphinxcode{\sphinxupquote{toSet}} when applied to a function, or \sphinxcode{\sphinxupquote{toSet}} when applied to a relation.
By referring to the set of tuples of a function \sphinxcode{\sphinxupquote{f}} indirectly by means of \sphinxcode{\sphinxupquote{toSet(f)}}, the set of tuples of a relation \sphinxcode{\sphinxupquote{R}} by means of \sphinxcode{\sphinxupquote{toSet(R)}}, or the relation corresponding to a function \sphinxcode{\sphinxupquote{g}} by \sphinxcode{\sphinxupquote{toRelation(g)}}, it is possible to use the declarative forms

\begin{sphinxVerbatim}[commandchars=\\\{\}]
\PYG{k+kt}{find} R : relation \PYG{k+kt}{of} \PYG{o}{(}\PYG{k+kt}{int}\PYG{o}{(}0..1\PYG{o}{)} \PYG{o}{*} \PYG{k+kt}{int}\PYG{o}{(}0..1\PYG{o}{)}\PYG{o}{)}
such that toSet\PYG{o}{(}R\PYG{o}{)} = \PYG{o}{\PYGZob{}}\PYG{o}{(}0,0\PYG{o}{)},\PYG{o}{(}0,1\PYG{o}{)},\PYG{o}{(}1,1\PYG{o}{)}\PYG{o}{\PYGZcb{}}

\PYG{k+kt}{find} f : function \PYG{k+kt}{int}\PYG{o}{(}0..1\PYG{o}{)} \PYGZhy{}\PYGZhy{}\PYGZgt{} \PYG{k+kt}{int}\PYG{o}{(}0..1\PYG{o}{)}
such that toSet\PYG{o}{(}f\PYG{o}{)} = \PYG{o}{\PYGZob{}}\PYG{o}{(}0,0\PYG{o}{)},\PYG{o}{(}1,1\PYG{o}{)}\PYG{o}{\PYGZcb{}}

\PYG{k+kt}{find} g : function \PYG{k+kt}{int}\PYG{o}{(}0..1\PYG{o}{)} \PYGZhy{}\PYGZhy{}\PYGZgt{} \PYG{k+kt}{int}\PYG{o}{(}0..1\PYG{o}{)}
such that toRelation\PYG{o}{(}g\PYG{o}{)} = relation\PYG{o}{(}\PYG{o}{(}0,0\PYG{o}{)},\PYG{o}{(}1,1\PYG{o}{)}\PYG{o}{)}
\end{sphinxVerbatim}

to indirectly recover the relation or function that corresponds to a set of tuples, or the function that corresponds to a relation.
This will fail to yield a solution if a function corresponding to a set of tuples or relation is sought, but that set of tuples or relation does not actually determine a function.
An error results if a relation corresponding to a set of tuples is sought, but not all tuples have the same number of elements.

\subsection{Function operators}
\label{\detokenize{essence:function-operators}}

\begin{savenotes}\sphinxattablestart
\centering
\begin{tabulary}{\linewidth}[t]{|T|T|}
\hline

\sphinxcode{\sphinxupquote{defined}}
&
set of values for which function is defined
\\
\hline
\sphinxcode{\sphinxupquote{image}}
&
\sphinxcode{\sphinxupquote{image(f,x)}} is the same as \sphinxcode{\sphinxupquote{f(x)}}
\\
\hline
\sphinxcode{\sphinxupquote{imageSet}}
&
\sphinxcode{\sphinxupquote{imageSet(f,x)}} is \sphinxcode{\sphinxupquote{\{f(x)\}}} if \sphinxcode{\sphinxupquote{f(x)}} is
defined, or empty if \sphinxcode{\sphinxupquote{f(x)}} is not defined
\\
\hline
\sphinxcode{\sphinxupquote{inverse}}
&
test if two functions are inverses of each other
\\
\hline
\sphinxcode{\sphinxupquote{preImage}}
&
set of elements mapped by function to an element
\\
\hline
\sphinxcode{\sphinxupquote{range}}
&
set of values of function
\\
\hline
\sphinxcode{\sphinxupquote{restrict}}
&
function restricted to a domain
\\
\hline
\end{tabulary}
\par
\sphinxattableend\end{savenotes}

Operators \sphinxcode{\sphinxupquote{defined}} and \sphinxcode{\sphinxupquote{range}} yield the sets of values that a function maps between.
For all functions \sphinxcode{\sphinxupquote{f}}, the set \sphinxcode{\sphinxupquote{toSet(f)}} is contained in the Cartesian product of sets \sphinxcode{\sphinxupquote{defined(f)}} and \sphinxcode{\sphinxupquote{range(f)}}.

For a function \sphinxcode{\sphinxupquote{f}} and a domain \sphinxcode{\sphinxupquote{D}}, the expression \sphinxcode{\sphinxupquote{restrict(f,D)}} denotes the function that is defined on the values in \sphinxcode{\sphinxupquote{D}} for which \sphinxcode{\sphinxupquote{f}} is defined, and that also coincides with \sphinxcode{\sphinxupquote{f}} where it is defined.

\begin{sphinxVerbatim}[commandchars=\\\{\}]
\PYG{k+kt}{letting} f \PYG{k+kt}{be} function\PYG{o}{(}0\PYGZhy{}\PYGZhy{}\PYGZgt{}1,3\PYGZhy{}\PYGZhy{}\PYGZgt{}4\PYG{o}{)}
\PYG{k+kt}{letting} D \PYG{k+kt}{be} \PYG{k+kt}{domain} \PYG{k+kt}{int}\PYG{o}{(}0,2\PYG{o}{)}
\PYG{k+kt}{find} g : function \PYG{k+kt}{int}\PYG{o}{(}0..4\PYG{o}{)}\PYGZhy{}\PYGZhy{}\PYGZgt{}int\PYG{o}{(}0..4\PYG{o}{)} such that
  g = restrict\PYG{o}{(}f, D\PYG{o}{)} \PYGZdl{} function\PYG{o}{(}0\PYGZhy{}\PYGZhy{}\PYGZgt{}1\PYG{o}{)}
\PYG{k+kt}{find} a : bool such that \PYGZdl{} true
  a = \PYG{o}{(} \PYG{o}{(}defined\PYG{o}{(}g\PYG{o}{)} = defined\PYG{o}{(}f\PYG{o}{)} intersect toSet\PYG{o}{(}[i \PYG{o}{\textbar{}} i : D]\PYG{o}{)}\PYG{o}{)}
    /\PYGZbs{} \PYG{o}{(}forAll x in defined\PYG{o}{(}g\PYG{o}{)} . g\PYG{o}{(}x\PYG{o}{)} = f\PYG{o}{(}x\PYG{o}{)}\PYG{o}{)} \PYG{o}{)}
\end{sphinxVerbatim}

Applying \sphinxcode{\sphinxupquote{image}} to values for which the function is not defined may lead to unintended unsatisfiability.
The Conjure specific \sphinxcode{\sphinxupquote{imageSet}} operator is useful for partial functions to avoid unsatisfiability in these cases.
The original Essence definition allows \sphinxcode{\sphinxupquote{image}} to represent the image of a function with respect to either an element or a set.
Conjure does not currently support taking the \sphinxcode{\sphinxupquote{image}} or \sphinxcode{\sphinxupquote{preImage}} of a function with respect to a set of elements.

The \sphinxcode{\sphinxupquote{inverse}} operator tests whether its function arguments are inverses of each other.

\begin{sphinxVerbatim}[commandchars=\\\{\}]
\PYG{k+kt}{find} a : bool such that a = inverse\PYG{o}{(}function\PYG{o}{(}0\PYGZhy{}\PYGZhy{}\PYGZgt{}1\PYG{o}{)},function\PYG{o}{(}1\PYGZhy{}\PYGZhy{}\PYGZgt{}0\PYG{o}{)}\PYG{o}{)} \PYGZdl{} true
\PYG{k+kt}{find} b : bool such that b = inverse\PYG{o}{(}function\PYG{o}{(}0\PYGZhy{}\PYGZhy{}\PYGZgt{}1\PYG{o}{)},function\PYG{o}{(}1\PYGZhy{}\PYGZhy{}\PYGZgt{}1\PYG{o}{)}\PYG{o}{)} \PYGZdl{} false
\end{sphinxVerbatim}

\subsection{Matrix operators}
\label{\detokenize{essence:matrix-operators}}
The following operator returns a matrix:

\begin{savenotes}\sphinxattablestart
\centering
\begin{tabulary}{\linewidth}[t]{|T|T|}
\hline

\sphinxcode{\sphinxupquote{flatten}}
&
list of entries from matrix
\\
\hline
\end{tabulary}
\par
\sphinxattableend\end{savenotes}

\sphinxcode{\sphinxupquote{flatten}} takes 1 or 2 arguments.
With one argument, \sphinxcode{\sphinxupquote{flatten}} returns a list containing the entries of a matrix with any number of dimensions, listed in the lexicographic order of the tuples of indices specifying each entry.
With two arguments \sphinxcode{\sphinxupquote{flatten(n,M)}}, the first argument \sphinxcode{\sphinxupquote{n}} is a constant integer that indicates the depth of flattening: the first \sphinxcode{\sphinxupquote{n+1}} dimensions are flattened into one dimension.
Note that \sphinxcode{\sphinxupquote{flatten(0,M) = M}} always holds.
The one-argument form works like an unbounded-depth flattening.

The following operators yield Boolean values:

\begin{savenotes}\sphinxattablestart
\centering
\begin{tabulary}{\linewidth}[t]{|T|T|}
\hline

\sphinxcode{\sphinxupquote{allDiff}}
&
test if all entries of a list are different
\\
\hline
\sphinxcode{\sphinxupquote{alldifferent\_except}}
&
test if all entries of a list differ,
possibly except value specified in second argument
\\
\hline
\end{tabulary}
\par
\sphinxattableend\end{savenotes}

The following illustrate \sphinxcode{\sphinxupquote{allDiff}} and \sphinxcode{\sphinxupquote{alldifferent\_except}}:

\begin{sphinxVerbatim}[commandchars=\\\{\}]
\PYG{k+kt}{find} a : bool such that a = allDiff\PYG{o}{(}[1,2,4,1]\PYG{o}{)} \PYGZdl{} false
\PYG{k+kt}{find} b : bool such that b = alldifferent\PYGZus{}except\PYG{o}{(}[1,2,4,1], 1\PYG{o}{)} \PYGZdl{} true
\end{sphinxVerbatim}

\subsection{Partition operators}
\label{\detokenize{essence:partition-operators}}

\begin{savenotes}\sphinxattablestart
\centering
\begin{tabulary}{\linewidth}[t]{|T|T|}
\hline

\sphinxcode{\sphinxupquote{apart}}
&
test if a list of elements are not all contained
in one part of the partition
\\
\hline
\sphinxcode{\sphinxupquote{participants}}
&
union of all parts of a partition
\\
\hline
\sphinxcode{\sphinxupquote{party}}
&
part of partition that contains specified element
\\
\hline
\sphinxcode{\sphinxupquote{parts}}
&
partition to its set of parts
\\
\hline
\sphinxcode{\sphinxupquote{together}}
&
test if a list of elements are all in the same
part of the partition
\\
\hline
\end{tabulary}
\par
\sphinxattableend\end{savenotes}

Examples:

\begin{sphinxVerbatim}[commandchars=\\\{\}]
\PYG{k+kt}{letting} P \PYG{k+kt}{be} partition\PYG{o}{(}\PYG{o}{\PYGZob{}}1,2\PYG{o}{\PYGZcb{}},\PYG{o}{\PYGZob{}}3\PYG{o}{\PYGZcb{}},\PYG{o}{\PYGZob{}}4,5,6\PYG{o}{\PYGZcb{}}\PYG{o}{)}
\PYG{k+kt}{find} a : bool such that a = apart\PYG{o}{(}\PYG{o}{\PYGZob{}}3,5\PYG{o}{\PYGZcb{}},P\PYG{o}{)} /\PYGZbs{} !together\PYG{o}{(}\PYG{o}{\PYGZob{}}1,2,5\PYG{o}{\PYGZcb{}},P\PYG{o}{)} \PYGZdl{} true
\PYG{k+kt}{find} b : \PYG{k+kt}{set} \PYG{k+kt}{of} \PYG{k+kt}{int}\PYG{o}{(}1..6\PYG{o}{)} such that b = participants\PYG{o}{(}P\PYG{o}{)} \PYGZdl{} \PYG{o}{\PYGZob{}}1,2,3,4,5,6\PYG{o}{\PYGZcb{}}
\PYG{k+kt}{find} c : \PYG{k+kt}{set} \PYG{k+kt}{of} \PYG{k+kt}{int}\PYG{o}{(}1..6\PYG{o}{)} such that c = party\PYG{o}{(}4,P\PYG{o}{)} \PYGZdl{} \PYG{o}{\PYGZob{}}4,5,6\PYG{o}{\PYGZcb{}}
\PYG{k+kt}{find} d : bool such that d = \PYG{o}{(}\PYG{o}{\PYGZob{}}\PYG{o}{\PYGZob{}}1,2\PYG{o}{\PYGZcb{}},\PYG{o}{\PYGZob{}}3\PYG{o}{\PYGZcb{}},\PYG{o}{\PYGZob{}}4,5,6\PYG{o}{\PYGZcb{}}\PYG{o}{\PYGZcb{}} = parts\PYG{o}{(}P\PYG{o}{)}\PYG{o}{)} \PYGZdl{} true
\PYG{k+kt}{find} e : bool such that e = \PYG{o}{(}together\PYG{o}{(}\PYG{o}{\PYGZob{}}1,7\PYG{o}{\PYGZcb{}},P\PYG{o}{)} \PYGZbs{}/ apart\PYG{o}{(}\PYG{o}{\PYGZob{}}1,7\PYG{o}{\PYGZcb{}},P\PYG{o}{)}\PYG{o}{)} \PYGZdl{} false
\end{sphinxVerbatim}

These semantics follow the original Essence definition.
In contrast, in older versions of Conjure the relationship
\begin{quote}

\begin{DUlineblock}{0em}
\item[] \sphinxcode{\sphinxupquote{apart(L,P) = !together(L,P)}}
\end{DUlineblock}
\end{quote}

held for all lists \sphinxcode{\sphinxupquote{L}} and partitions \sphinxcode{\sphinxupquote{P}}.

\subsection{List combining operators}
\label{\detokenize{essence:list-combining-operators}}
Each of the operators
\begin{quote}

\begin{DUlineblock}{0em}
\item[] \sphinxcode{\sphinxupquote{sum    product    and    or    xor}}
\end{DUlineblock}
\end{quote}

applies an associative combining operator to elements of a list or set.
A list may also be given as a comprehension that specifies the elements of a set or domain that satisfy some conditions.

The following relationships hold for all integers \sphinxcode{\sphinxupquote{x}} and \sphinxcode{\sphinxupquote{y}}:
\begin{quote}

\begin{DUlineblock}{0em}
\item[] \sphinxcode{\sphinxupquote{sum({[}x,y{]}) = (x + y)}}
\item[] \sphinxcode{\sphinxupquote{product({[}x,y{]}) = (x * y)}}
\end{DUlineblock}
\end{quote}

The following relationships hold for all Booleans \sphinxcode{\sphinxupquote{a}} and \sphinxcode{\sphinxupquote{b}}:
\begin{quote}

\begin{DUlineblock}{0em}
\item[] \sphinxcode{\sphinxupquote{and({[}a,b{]}) = (a /\textbackslash{} b)}}
\item[] \sphinxcode{\sphinxupquote{or({[}a,b{]}) = (a \textbackslash{}/ b)}}
\item[] \sphinxcode{\sphinxupquote{xor({[}a,b{]}) = ((a \textbackslash{}/ b) /\textbackslash{} !(a /\textbackslash{} b))}}
\end{DUlineblock}
\end{quote}

Examples:

\begin{sphinxVerbatim}[commandchars=\\\{\}]
\PYG{k+kt}{find} x : \PYG{k+kt}{int}\PYG{o}{(}0..9\PYG{o}{)} such that x = sum\PYG{o}{(} \PYG{o}{\PYGZob{}}1,2,3\PYG{o}{\PYGZcb{}} \PYG{o}{)} \PYGZdl{} 6
\PYG{k+kt}{find} y : \PYG{k+kt}{int}\PYG{o}{(}0..9\PYG{o}{)} such that y = product\PYG{o}{(} [1,2,4] \PYG{o}{)} \PYGZdl{} 8
\PYG{k+kt}{find} a : bool such that a = and\PYG{o}{(}[xor\PYG{o}{(}[true,false]\PYG{o}{)},or\PYG{o}{(}[false,true]\PYG{o}{)}]\PYG{o}{)} \PYGZdl{} true
\end{sphinxVerbatim}

Quantification over a finite set or finite domain of values is supported by \sphinxcode{\sphinxupquote{forAll}} and \sphinxcode{\sphinxupquote{exists}}.
These quantifiers yield Boolean values and are internally treated as \sphinxcode{\sphinxupquote{and}} and \sphinxcode{\sphinxupquote{or}}, respectively, applied to the lists of values corresponding to the set or domain.
The following snippets illustrate the use of quantifiers.

\begin{sphinxVerbatim}[commandchars=\\\{\}]
\PYG{k+kt}{find} a : bool such that a = forAll i in \PYG{o}{\PYGZob{}}0,1,2\PYG{o}{\PYGZcb{}} . i=i*i \PYGZdl{} false
\PYG{k+kt}{find} b : bool such that b = exists i : \PYG{k+kt}{int}\PYG{o}{(}0..4\PYG{o}{)} . i*i=i \PYGZdl{} true
\end{sphinxVerbatim}

The same variable can be reused for multiple quantifications, as a quantified variable has scope that is local to its quantifier.
However, avoid using the same name both for quantification and as a global decision variable in a \sphinxcode{\sphinxupquote{find}}, as this is treated as an error by Savile Row.

An alternative quantifier-like syntax
\begin{quote}

\begin{DUlineblock}{0em}
\item[] \sphinxcode{\sphinxupquote{sum i in I . f(i)}}
\end{DUlineblock}
\end{quote}

is supported for the \sphinxcode{\sphinxupquote{sum}} and \sphinxcode{\sphinxupquote{product}} operators.

\subsection{Comprehensions}
\label{\detokenize{essence:comprehensions}}
A list can be constructed by means of a comprehension.
A list comprehension is declared by using the usual square brackets \sphinxcode{\sphinxupquote{{[}}} and \sphinxcode{\sphinxupquote{{]}}}, inside which is a generator expression possibly involving some parameter variables, followed by \sphinxcode{\sphinxupquote{\textbar{}}}, followed by a comma (\sphinxcode{\sphinxupquote{,}}) separated sequence of conditions defining the values that all the parameter variables may take, or Boolean expressions.
The value of a list comprehension is a list containing all the values of the generator expression corresponding to those values of the parameter variables for which all the Boolean expressions evaluate to \sphinxcode{\sphinxupquote{true}}.

In a Boolean expression controlling a comprehension, if \sphinxcode{\sphinxupquote{L}} is a list then \sphinxcode{\sphinxupquote{v \textless{}- L}} behaves similarly to how the expression \sphinxcode{\sphinxupquote{v in toMSet(L)}} is treated in a quantification.

Examples of list comprehensions:

\begin{sphinxVerbatim}[commandchars=\\\{\}]
\PYG{k+kt}{find} x : \PYG{k+kt}{int}\PYG{o}{(}0..999\PYG{o}{)} such that x = product\PYG{o}{(} [i\PYGZhy{}1 \PYG{o}{\textbar{}} i \PYGZlt{}\PYGZhy{} [5,6,7]] \PYG{o}{)} \PYGZdl{} 120
\PYG{k+kt}{letting} M \PYG{k+kt}{be} [1,0,0,1,0]
\PYG{k+kt}{letting} I \PYG{k+kt}{be} \PYG{k+kt}{domain} \PYG{k+kt}{int}\PYG{o}{(}1..5\PYG{o}{)}
\PYG{k+kt}{find} y : \PYG{k+kt}{int}\PYG{o}{(}0..9\PYG{o}{)} such that y = sum\PYG{o}{(} [toInt\PYG{o}{(}\PYG{o}{(}i=j\PYG{o}{)} /\PYGZbs{} \PYG{o}{(}M[j]\PYGZgt{}0\PYG{o}{)}\PYG{o}{)} \PYG{o}{\textbar{}} i : I, j \PYGZlt{}\PYGZhy{} M] \PYG{o}{)} \PYGZdl{} 2
\PYG{k+kt}{find} a : bool such that a = and\PYG{o}{(}[u\PYGZlt{}v \PYG{o}{\textbar{}} \PYG{o}{(}u,v\PYG{o}{)} \PYGZlt{}\PYGZhy{} [\PYG{o}{(}0,1\PYG{o}{)},\PYG{o}{(}2**10,2**11\PYG{o}{)},\PYG{o}{(}\PYGZhy{}1,1\PYG{o}{)}] ]\PYG{o}{)} \PYGZdl{} true
\end{sphinxVerbatim}

\chapter{Demonstrations}
\label{\detokenize{demonstrations:demonstrations}}\label{\detokenize{demonstrations:id1}}\label{\detokenize{demonstrations::doc}}
We demonstrate the use of Conjure for some small problems.

\section{Number puzzle}
\label{\detokenize{demonstrations:number-puzzle}}
We first show how to solve a classic \sphinxhref{https://en.wikipedia.org/wiki/Verbal\_arithmetic}{word addition} puzzle, due to Dudeney \sphinxcite{zreferences:dudeney1924puzzle}.
This is a small toy example, but already illustrates some interesting features of Conjure.

\begin{sphinxVerbatim}[commandchars=\\\{\}]
   \PYG{n}{SEND}
\PYG{o}{+}  \PYG{n}{MORE}
\PYG{o}{\PYGZhy{}}\PYG{o}{\PYGZhy{}}\PYG{o}{\PYGZhy{}}\PYG{o}{\PYGZhy{}}\PYG{o}{\PYGZhy{}}\PYG{o}{\PYGZhy{}}\PYG{o}{\PYGZhy{}}
\PYG{o}{=} \PYG{n}{MONEY}
\end{sphinxVerbatim}

Here each letter represents a numeric digit in an addition, and we are asked to find an assignment of digits to letters so that the number represented by the digits SEND when added to the number represented by MORE yields the number represented by MONEY.

\subsection{Initial model}
\label{\detokenize{demonstrations:initial-model}}
We are looking for a mapping (a function) from letters to digits.
We can represent the different letters as an enumerated type, with each letter appearing only once.
We then need to express what it means for the digits of the sum to behave as we expect; a natural approach is to introduce carry digits and use those to express the sum digit-wise.
There are only at most two digits and a carry digit being added at each step, so the carry digits cannot be larger than 2.

\begin{sphinxVerbatim}[commandchars=\\\{\}]
language Essence 1.3
\PYG{k+kt}{letting} letters \PYG{k+kt}{be} new type enum \PYG{o}{\PYGZob{}}S,E,N,D,M,O,R,Y\PYG{o}{\PYGZcb{}}
\PYG{k+kt}{find} f : function letters \PYGZhy{}\PYGZhy{}\PYGZgt{} \PYG{k+kt}{int}\PYG{o}{(}0..9\PYG{o}{)}
\PYG{k+kt}{find} carry1,carry2,carry3,carry4 : \PYG{k+kt}{int}\PYG{o}{(}0..2\PYG{o}{)}
such that
           f\PYG{o}{(}D\PYG{o}{)} + f\PYG{o}{(}E\PYG{o}{)} = f\PYG{o}{(}Y\PYG{o}{)} + 10*carry1,
  carry1 + f\PYG{o}{(}N\PYG{o}{)} + f\PYG{o}{(}R\PYG{o}{)} = f\PYG{o}{(}E\PYG{o}{)} + 10*carry2,
  carry2 + f\PYG{o}{(}E\PYG{o}{)} + f\PYG{o}{(}O\PYG{o}{)} = f\PYG{o}{(}N\PYG{o}{)} + 10*carry3,
  carry3 + f\PYG{o}{(}S\PYG{o}{)} + f\PYG{o}{(}M\PYG{o}{)} = f\PYG{o}{(}O\PYG{o}{)} + 10*carry4,
  carry4 = f\PYG{o}{(}M\PYG{o}{)}
\end{sphinxVerbatim}

Each Essence specification can optionally contain a declaration of which dialect of Essence it is written in.
The current version of Essence is 1.3.
We leave out this declaration in the remaining examples to avoid repetition.

This model is stored in \sphinxcode{\sphinxupquote{sm1.essence}}; let’s use Conjure to find the solution:

\begin{sphinxVerbatim}[commandchars=\\\{\}]
conjure solve \PYGZhy{}ac sm1.essence
\end{sphinxVerbatim}

Unless we specify what to call the solution, it is saved as \sphinxcode{\sphinxupquote{sm1.solution}}.

\begin{sphinxVerbatim}[commandchars=\\\{\}]
\PYG{k+kt}{letting} carry1 \PYG{k+kt}{be} 0
\PYG{k+kt}{letting} carry2 \PYG{k+kt}{be} 0
\PYG{k+kt}{letting} carry3 \PYG{k+kt}{be} 0
\PYG{k+kt}{letting} carry4 \PYG{k+kt}{be} 0
\PYG{k+kt}{letting} f \PYG{k+kt}{be} function\PYG{o}{(}S \PYGZhy{}\PYGZhy{}\PYGZgt{} 0, E \PYGZhy{}\PYGZhy{}\PYGZgt{} 0, N \PYGZhy{}\PYGZhy{}\PYGZgt{} 0, D \PYGZhy{}\PYGZhy{}\PYGZgt{} 0, M \PYGZhy{}\PYGZhy{}\PYGZgt{} 0,
  O \PYGZhy{}\PYGZhy{}\PYGZgt{} 0, R \PYGZhy{}\PYGZhy{}\PYGZgt{} 0, Y \PYGZhy{}\PYGZhy{}\PYGZgt{} 0\PYG{o}{)}
\end{sphinxVerbatim}

This is clearly not what we wanted.
We haven’t specified all the constraints in the problem!

\subsection{Identifying a missing constraint}
\label{\detokenize{demonstrations:identifying-a-missing-constraint}}
In these kinds of puzzles, usually we need each letter to map to a different digit: we need an injective function.
Let’s replace the line

\begin{sphinxVerbatim}[commandchars=\\\{\}]
\PYG{k+kt}{find} f : function letters \PYGZhy{}\PYGZhy{}\PYGZgt{} \PYG{k+kt}{int}\PYG{o}{(}0..9\PYG{o}{)}
\end{sphinxVerbatim}

by

\begin{sphinxVerbatim}[commandchars=\\\{\}]
\PYG{k+kt}{find} f : function \PYG{o}{(}injective\PYG{o}{)} letters \PYGZhy{}\PYGZhy{}\PYGZgt{} \PYG{k+kt}{int}\PYG{o}{(}0..9\PYG{o}{)}
\end{sphinxVerbatim}

and save the result in file \sphinxcode{\sphinxupquote{sm2.essence}}.
Now let’s run Conjure again on the new model:

\begin{sphinxVerbatim}[commandchars=\\\{\}]
conjure solve \PYGZhy{}ac sm2.essence
\end{sphinxVerbatim}

This time the solution \sphinxcode{\sphinxupquote{sm2.solution}} looks more like what we wanted:

\begin{sphinxVerbatim}[commandchars=\\\{\}]
letting carry1 be \PYG{l+m}{1}
letting carry2 be \PYG{l+m}{0}
letting carry3 be \PYG{l+m}{1}
letting carry4 be \PYG{l+m}{0}
letting f be \PYG{k}{function}\PYG{o}{(}S \PYGZhy{}\PYGZhy{}\PYGZgt{} \PYG{l+m}{2}, E \PYGZhy{}\PYGZhy{}\PYGZgt{} \PYG{l+m}{8}, N \PYGZhy{}\PYGZhy{}\PYGZgt{} \PYG{l+m}{1}, D \PYGZhy{}\PYGZhy{}\PYGZgt{} \PYG{l+m}{7}, M \PYGZhy{}\PYGZhy{}\PYGZgt{} \PYG{l+m}{0},
  O \PYGZhy{}\PYGZhy{}\PYGZgt{} \PYG{l+m}{3}, R \PYGZhy{}\PYGZhy{}\PYGZgt{} \PYG{l+m}{6}, Y \PYGZhy{}\PYGZhy{}\PYGZgt{} \PYG{l+m}{5}\PYG{o}{)}
\end{sphinxVerbatim}

\subsection{Final model}
\label{\detokenize{demonstrations:final-model}}
There is still something strange with \sphinxcode{\sphinxupquote{sm2.essence}}.
We usually do not allow a number to begin with a zero digit, but the solution maps M to 0.
Let’s add the missing constraints to file \sphinxcode{\sphinxupquote{sm3.essence}}:

\begin{sphinxVerbatim}[commandchars=\\\{\}]
\PYG{k+kt}{letting} letters \PYG{k+kt}{be} new type enum \PYG{o}{\PYGZob{}}S,E,N,D,M,O,R,Y\PYG{o}{\PYGZcb{}}
\PYG{k+kt}{find} f : function \PYG{o}{(}injective\PYG{o}{)} letters \PYGZhy{}\PYGZhy{}\PYGZgt{} \PYG{k+kt}{int}\PYG{o}{(}0..9\PYG{o}{)}
\PYG{k+kt}{find} carry1,carry2,carry3,carry4 : \PYG{k+kt}{int}\PYG{o}{(}0..2\PYG{o}{)}
such that
           f\PYG{o}{(}D\PYG{o}{)} + f\PYG{o}{(}E\PYG{o}{)} = f\PYG{o}{(}Y\PYG{o}{)} + 10*carry1,
  carry1 + f\PYG{o}{(}N\PYG{o}{)} + f\PYG{o}{(}R\PYG{o}{)} = f\PYG{o}{(}E\PYG{o}{)} + 10*carry2,
  carry2 + f\PYG{o}{(}E\PYG{o}{)} + f\PYG{o}{(}O\PYG{o}{)} = f\PYG{o}{(}N\PYG{o}{)} + 10*carry3,
  carry3 + f\PYG{o}{(}S\PYG{o}{)} + f\PYG{o}{(}M\PYG{o}{)} = f\PYG{o}{(}O\PYG{o}{)} + 10*carry4,
  carry4 = f\PYG{o}{(}M\PYG{o}{)},
  M \PYGZgt{} 0, S \PYGZgt{} 0
\end{sphinxVerbatim}

Let’s try again:

\begin{sphinxVerbatim}[commandchars=\\\{\}]
conjure solve \PYGZhy{}ac sm3.essence
\end{sphinxVerbatim}

This now leads to the solution we expected:

\begin{sphinxVerbatim}[commandchars=\\\{\}]
\PYG{k+kt}{letting} carry1 \PYG{k+kt}{be} 1
\PYG{k+kt}{letting} carry2 \PYG{k+kt}{be} 1
\PYG{k+kt}{letting} carry3 \PYG{k+kt}{be} 0
\PYG{k+kt}{letting} carry4 \PYG{k+kt}{be} 1
\PYG{k+kt}{letting} f \PYG{k+kt}{be} function\PYG{o}{(}S \PYGZhy{}\PYGZhy{}\PYGZgt{} 9, E \PYGZhy{}\PYGZhy{}\PYGZgt{} 5, N \PYGZhy{}\PYGZhy{}\PYGZgt{} 6, D \PYGZhy{}\PYGZhy{}\PYGZgt{} 7, M \PYGZhy{}\PYGZhy{}\PYGZgt{} 1,
  O \PYGZhy{}\PYGZhy{}\PYGZgt{} 0, R \PYGZhy{}\PYGZhy{}\PYGZgt{} 8, Y \PYGZhy{}\PYGZhy{}\PYGZgt{} 2\PYG{o}{)}
\end{sphinxVerbatim}

Note that the solution includes both the mapping we were looking for, as well as values for the carry digits that were introduced to express the constraints.

Finally, let’s check that there are no more solutions:

\begin{sphinxVerbatim}[commandchars=\\\{\}]
conjure solve \PYGZhy{}ac sm3.essence \PYGZhy{}\PYGZhy{}number\PYGZhy{}of\PYGZhy{}solutions\PYG{o}{=}all
\end{sphinxVerbatim}

This confirms that there is indeed only one solution.
As an exercise, verify that the first two models have multiple solutions, and that the solution given by the third model is among these.
(The first has 1155 solutions, the second 25.)

\section{Labelled connected graphs}
\label{\detokenize{demonstrations:labelled-connected-graphs}}
We now illustrate the use of Conjure for a more realistic modelling task, to enumerate all labelled connected graphs.
The number of labelled connected graphs over a fixed set of n distinct labels grows quickly; this is \sphinxhref{http://oeis.org/A001187}{OEIS sequence A001187}.

We first need to decide how to represent graphs.
A standard representation is to list the edges.
One natural representation for each edge is as a set of two distinct vertices.
Vertices of the graph are labelled with integers between 1 and n, and each vertex is regarded as part of the graph, whether there is some edge involving that vertex or not.

\begin{sphinxVerbatim}[commandchars=\\\{\}]
\PYG{k+kt}{letting} n \PYG{k+kt}{be} 4
\PYG{k+kt}{letting} G \PYG{k+kt}{be} \PYG{o}{\PYGZob{}}\PYG{o}{\PYGZob{}}1,2\PYG{o}{\PYGZcb{}},\PYG{o}{\PYGZob{}}2,3\PYG{o}{\PYGZcb{}},\PYG{o}{\PYGZob{}}3,4\PYG{o}{\PYGZcb{}}\PYG{o}{\PYGZcb{}}
\end{sphinxVerbatim}

In this specification, we declare two aliases.
The number of vertices n is first defined as 4.
Then G is defined as a set of edges.

This specification is saved in a file \sphinxcode{\sphinxupquote{path-4.param}} that we refer to later.
We should also have a different graph that is not connected:

\begin{sphinxVerbatim}[commandchars=\\\{\}]
\PYG{k+kt}{letting} n \PYG{k+kt}{be} 4
\PYG{k+kt}{letting} G \PYG{k+kt}{be} \PYG{o}{\PYGZob{}}\PYG{o}{\PYGZob{}}1,2\PYG{o}{\PYGZcb{}},\PYG{o}{\PYGZob{}}4,3\PYG{o}{\PYGZcb{}}\PYG{o}{\PYGZcb{}}
\end{sphinxVerbatim}

which is saved in file \sphinxcode{\sphinxupquote{disconnected-4.param}}.

We now need to express what it means for a graph to be connected.

\subsection{Model 1: distance matrix}
\label{\detokenize{demonstrations:model-1-distance-matrix}}
In our first attempt, we use a matrix of distances.
Each entry \sphinxcode{\sphinxupquote{reach{[}u,v{]}}} represents the length of a shortest path from u to v, or n if there is no path from u to v.
To enforce this property, we use several constraints, one for each possible length; there are four ranges of values we need to cover.
A distance of 0 happens when u and v are the same vertex.
A distance of 1 happens when there is an edge from u to v.
When the distance is greater than 1 but less than n, then there must be some vertex that is a neighbour of u from which v is reachable in one less step.
Finally, the distance of n is used when no neighbour of u can reach v (and in this case, the neighbours all have distance of n to v as well).

\begin{sphinxVerbatim}[commandchars=\\\{\}]
\PYG{k+kt}{given} n : \PYG{k+kt}{int}\PYG{o}{(}1..\PYG{o}{)}
\PYG{k+kt}{letting} vertices \PYG{k+kt}{be} \PYG{k+kt}{domain} \PYG{k+kt}{int}\PYG{o}{(}1..n\PYG{o}{)}
\PYG{k+kt}{given} G : \PYG{k+kt}{set} \PYG{k+kt}{of} \PYG{k+kt}{set} \PYG{o}{(}size 2\PYG{o}{)} \PYG{k+kt}{of} vertices
\PYG{k+kt}{find} reach : matrix indexed by [vertices, vertices] \PYG{k+kt}{of} \PYG{k+kt}{int}\PYG{o}{(}0..n\PYG{o}{)}
such that
  forAll u,v : vertices .
     \PYG{o}{(}\PYG{o}{(}reach[u,v] = 0\PYG{o}{)} \PYGZhy{}\PYGZgt{} \PYG{o}{(}u=v\PYG{o}{)}\PYG{o}{)}
  /\PYGZbs{} \PYG{o}{(}\PYG{o}{(}reach[u,v] = 1\PYG{o}{)} \PYGZhy{}\PYGZgt{} \PYG{o}{(}\PYG{o}{\PYGZob{}}u,v\PYG{o}{\PYGZcb{}} in G\PYG{o}{)}\PYG{o}{)}
  /\PYGZbs{} \PYG{o}{(}\PYG{o}{(}\PYG{o}{(}reach[u,v] \PYGZgt{} 1\PYG{o}{)} \PYGZbs{}/ \PYG{o}{(}reach[u,v] \PYGZlt{} n\PYG{o}{)}\PYG{o}{)} \PYGZhy{}\PYGZgt{}
      \PYG{o}{(}exists w : vertices . \PYG{o}{(}\PYG{o}{\PYGZob{}}u,w\PYG{o}{\PYGZcb{}} in G\PYG{o}{)} /\PYGZbs{} \PYG{o}{(}reach[w,v] = reach[u,v] \PYGZhy{} 1\PYG{o}{)}\PYG{o}{)}\PYG{o}{)}
  /\PYGZbs{} \PYG{o}{(}\PYG{o}{(}reach[u,v] = n\PYG{o}{)} \PYGZhy{}\PYGZgt{} \PYG{o}{(}forAll w : vertices . !\PYG{o}{(}\PYG{o}{\PYGZob{}}u,w\PYG{o}{\PYGZcb{}} in G\PYG{o}{)} \PYGZbs{}/ \PYG{o}{(}reach[w,v] = n\PYG{o}{)}\PYG{o}{)}\PYG{o}{)}
\PYG{k+kt}{find} connected : bool
such that
  connected = \PYG{o}{(}forAll u,v : vertices . reach[u,v] \PYGZlt{} n\PYG{o}{)}
\end{sphinxVerbatim}

This is stored in file \sphinxcode{\sphinxupquote{gc1.essence}}.
The values of n and G will be specified later as parameters, such as via the \sphinxcode{\sphinxupquote{path-4.param}} or \sphinxcode{\sphinxupquote{disconnected-4.param}} files.

In the model, first the matrix \sphinxcode{\sphinxupquote{reach}} is specified by imposing the four conditions that we mentioned.
Finally a Boolean variable is used to conveniently indicate whether the \sphinxcode{\sphinxupquote{reach}} matrix represents a connected graph or not; in a connected graph every vertex is reachable from every other vertex.

Let’s now try this model with the two graphs defined so far.

\begin{sphinxVerbatim}[commandchars=\\\{\}]
conjure solve \PYGZhy{}ac gc1.essence path\PYGZhy{}4.param
conjure solve \PYGZhy{}ac gc1.essence disconnected\PYGZhy{}4.param
\end{sphinxVerbatim}

In the solutions found by Conjure, the matrix \sphinxcode{\sphinxupquote{reach}} indicates the distances between each pair of vertices.
In the solution for the connected graph \sphinxcode{\sphinxupquote{gc1-path-4.solution}} all entries are at most 3.

\begin{sphinxVerbatim}[commandchars=\\\{\}]
\PYG{k+kt}{letting} connected \PYG{k+kt}{be} true
\PYG{k+kt}{letting} reach \PYG{k+kt}{be}
  [[0, 1, 2, 3; \PYG{k+kt}{int}\PYG{o}{(}1..4\PYG{o}{)}], [1, 0, 1, 2; \PYG{k+kt}{int}\PYG{o}{(}1..4\PYG{o}{)}],
   [2, 1, 0, 1; \PYG{k+kt}{int}\PYG{o}{(}1..4\PYG{o}{)}], [3, 2, 1, 0; \PYG{k+kt}{int}\PYG{o}{(}1..4\PYG{o}{)}]; \PYG{k+kt}{int}\PYG{o}{(}1..4\PYG{o}{)}]
\PYGZdl{} Visualisation for reach
\PYGZdl{} 0 1 2 3
\PYGZdl{} 1 0 1 2
\PYGZdl{} 2 1 0 1
\PYGZdl{} 3 2 1 0
\end{sphinxVerbatim}

In contrast, in the solution for the disconnected graph \sphinxcode{\sphinxupquote{gc1-disconnected-4.solution}} there are some entries that are 4:

\begin{sphinxVerbatim}[commandchars=\\\{\}]
\PYG{k+kt}{letting} connected \PYG{k+kt}{be} false
\PYG{k+kt}{letting} reach \PYG{k+kt}{be}
  [[0, 1, 4, 4; \PYG{k+kt}{int}\PYG{o}{(}1..4\PYG{o}{)}], [1, 0, 4, 4; \PYG{k+kt}{int}\PYG{o}{(}1..4\PYG{o}{)}],
   [4, 4, 0, 1; \PYG{k+kt}{int}\PYG{o}{(}1..4\PYG{o}{)}], [4, 4, 1, 0; \PYG{k+kt}{int}\PYG{o}{(}1..4\PYG{o}{)}]; \PYG{k+kt}{int}\PYG{o}{(}1..4\PYG{o}{)}]
\PYGZdl{} Visualisation for reach
\PYGZdl{} 0 1 4 4
\PYGZdl{} 1 0 4 4
\PYGZdl{} 4 4 0 1
\PYGZdl{} 4 4 1 0
\end{sphinxVerbatim}

Graphs with four vertices are good for quick testing but are too small to notice much difference between models.
Small differences are important for tasks such as enumerating many objects, when even a small difference is multiplied by the number of objects.
For testing we can create other parameter files containing graphs with more vertices.
Notice that we do not have to change the model, only the parameter files containing the input data.

Testing with larger graphs of say 1000 vertices, it becomes clear that this first model works but does not scale well.
It computes the lengths of the shortest paths between pairs of vertices, from which we can deduce whether the graph is connected.
This is quite round-about!
We can now try to improve the model by asking the system to do less work.
After all, we don’t actually need all the pairwise distances.

\subsection{Model 2: reachability matrix}
\label{\detokenize{demonstrations:model-2-reachability-matrix}}
In the following model, stored as file \sphinxcode{\sphinxupquote{gc2.essence}}, the reachability matrix uses Boolean values for the distances rather than integers, with \sphinxcode{\sphinxupquote{true}} representing reachable and \sphinxcode{\sphinxupquote{false}} unreachable.
Each entry \sphinxcode{\sphinxupquote{reach{[}u,v{]}}} represents whether it is possible to reach v by some path that starts at u.
This is modelled as the disjunction of three conditions: u is reachable from itself, any neighbour of u is reachable from it, and if v is not a neighbour of u then there should be a neighbour w of u so that v is reachable from w.

\begin{sphinxVerbatim}[commandchars=\\\{\}]
\PYG{k+kt}{given} n : \PYG{k+kt}{int}\PYG{o}{(}1..\PYG{o}{)}
\PYG{k+kt}{letting} vertices \PYG{k+kt}{be} \PYG{k+kt}{domain} \PYG{k+kt}{int}\PYG{o}{(}1..n\PYG{o}{)}
\PYG{k+kt}{given} G : \PYG{k+kt}{set} \PYG{k+kt}{of} \PYG{k+kt}{set} \PYG{o}{(}size 2\PYG{o}{)} \PYG{k+kt}{of} vertices
\PYG{k+kt}{find} reach : matrix indexed by [vertices, vertices] \PYG{k+kt}{of} bool
such that
  forAll u,v : vertices . reach[u,v] =
    \PYG{o}{(}\PYG{o}{(}u = v\PYG{o}{)} \PYGZbs{}/ \PYG{o}{(}\PYG{o}{\PYGZob{}}u,v\PYG{o}{\PYGZcb{}} in G\PYG{o}{)} \PYGZbs{}/
    \PYG{o}{(}exists w : vertices . \PYG{o}{(}\PYG{o}{\PYGZob{}}u,w\PYG{o}{\PYGZcb{}} in G\PYG{o}{)} /\PYGZbs{} reach[w,v]\PYG{o}{)}\PYG{o}{)}
\PYG{k+kt}{find} connected : bool
such that
  connected = \PYG{o}{(}forAll u,v : vertices . reach[u,v]\PYG{o}{)}
\end{sphinxVerbatim}

In the solutions found by Conjure, the reachability matrix contains regions of true entries indicating the connected components.

In the connected graph all entries are true:

\begin{sphinxVerbatim}[commandchars=\\\{\}]
\PYG{k+kt}{letting} connected \PYG{k+kt}{be} true
\PYG{k+kt}{letting} reach \PYG{k+kt}{be}
  [[true, true, true, true; \PYG{k+kt}{int}\PYG{o}{(}1..4\PYG{o}{)}], [true, true, true, true; \PYG{k+kt}{int}\PYG{o}{(}1..4\PYG{o}{)}],
   [true, true, true, true; \PYG{k+kt}{int}\PYG{o}{(}1..4\PYG{o}{)}], [true, true, true, true; \PYG{k+kt}{int}\PYG{o}{(}1..4\PYG{o}{)}];
   \PYG{k+kt}{int}\PYG{o}{(}1..4\PYG{o}{)}]
\PYGZdl{} Visualisation for reach
\PYGZdl{} T T T T
\PYGZdl{} T T T T
\PYGZdl{} T T T T
\PYGZdl{} T T T T
\end{sphinxVerbatim}

In contrast, in the disconnected graph there are some false entries:

\begin{sphinxVerbatim}[commandchars=\\\{\}]
\PYG{k+kt}{letting} connected \PYG{k+kt}{be} false
\PYG{k+kt}{letting} reach \PYG{k+kt}{be}
  [[true, true, false, false; \PYG{k+kt}{int}\PYG{o}{(}1..4\PYG{o}{)}], [true, true, false, false; \PYG{k+kt}{int}\PYG{o}{(}1..4\PYG{o}{)}],
   [false, false, true, true; \PYG{k+kt}{int}\PYG{o}{(}1..4\PYG{o}{)}], [false, false, true, true; \PYG{k+kt}{int}\PYG{o}{(}1..4\PYG{o}{)}];
   \PYG{k+kt}{int}\PYG{o}{(}1..4\PYG{o}{)}]
\PYGZdl{} Visualisation for reach
\PYGZdl{} T T \PYGZus{} \PYGZus{}
\PYGZdl{} T T \PYGZus{} \PYGZus{}
\PYGZdl{} \PYGZus{} \PYGZus{} T T
\PYGZdl{} \PYGZus{} \PYGZus{} T T
\end{sphinxVerbatim}

This model takes about half as long as the previous one, but is still rather slow for large graphs.

\subsection{Model 3: structured reachability matrices}
\label{\detokenize{demonstrations:model-3-structured-reachability-matrices}}
In the previous two models the solver may spend a long time early in the search process looking for ways to reach vertices that are far away, even though it would be more efficient to focus the early stages of search on vertices close by.
It is possible to improve performance by guiding the search to consider nearby vertices before vertices that are far from each other.
The following model \sphinxcode{\sphinxupquote{gc3.essence}} uses additional decision variables to more precisely control how the desired reachability matrix should be computed.
There are multiple reachability matrices.
Each corresponds to a specific maximum distance.
The first n by n matrix \sphinxcode{\sphinxupquote{reach{[}0{]}}} expresses reachability in one step, and is simply the adjacency matrix of the graph.
The entry \sphinxcode{\sphinxupquote{reach{[}k,u,v{]}}} expresses whether v is reachable from u via a path of length at most 2**k.
If a vertex v is reachable from some vertex u, then it can be reached in at most n-1 steps.
(Note: in this model a vertex cannot reach itself in zero steps, so a graph with a single vertex is not regarded as connected.)

\begin{sphinxVerbatim}[commandchars=\\\{\}]
\PYG{k+kt}{given} n : \PYG{k+kt}{int}\PYG{o}{(}1..\PYG{o}{)}
\PYG{k+kt}{letting} vertices \PYG{k+kt}{be} \PYG{k+kt}{domain} \PYG{k+kt}{int}\PYG{o}{(}1..n\PYG{o}{)}
\PYG{k+kt}{given} G : \PYG{k+kt}{set} \PYG{k+kt}{of} \PYG{k+kt}{set} \PYG{o}{(}size 2\PYG{o}{)} \PYG{k+kt}{of} vertices
\PYG{k+kt}{letting} m \PYG{k+kt}{be} sum\PYG{o}{(}[1 \PYG{o}{\textbar{}} i : \PYG{k+kt}{int}\PYG{o}{(}0..64\PYG{o}{)}, 2**i \PYGZlt{}= n]\PYG{o}{)}
\PYG{k+kt}{find} reach : matrix indexed by [int\PYG{o}{(}0..m\PYG{o}{)}, vertices, vertices] \PYG{k+kt}{of} bool
such that
  forAll u,v : vertices . reach[0,u,v] = \PYG{o}{(}\PYG{o}{\PYGZob{}}u,v\PYG{o}{\PYGZcb{}} in G\PYG{o}{)},
  forAll i : \PYG{k+kt}{int}\PYG{o}{(}0..\PYG{o}{(}m\PYGZhy{}1\PYG{o}{)}\PYG{o}{)} . forAll u,v : vertices . reach[i+1,u,v] =
    \PYG{o}{(}reach[i,u,v] \PYGZbs{}/ \PYG{o}{(}exists w : vertices . \PYG{o}{(}reach[i,u,w] /\PYGZbs{} reach[i,w,v]\PYG{o}{)}\PYG{o}{)}\PYG{o}{)},
\PYG{k+kt}{find} connected : bool
such that
  connected = \PYG{o}{(}forAll u,v : vertices . reach[m,u,v]\PYG{o}{)}
\end{sphinxVerbatim}

The variable m is used to compute the number of matrices that are required; this is the smallest integer that is not less than the base-2 logarithm of n.
(This is computed by discrete integration as Conjure currently does not support a logarithm operator; this may change in a future release.)
The value of \sphinxcode{\sphinxupquote{connected}} is then based on whether whether \sphinxcode{\sphinxupquote{reach{[}m{]}}} contains any false entries.

This model is the fastest yet, but it generates intermediate distance matrices, each containing n**2 variables.
We omit the solutions here, but they show how the number of true values increases, until reaching a fixed point.

\subsection{Model 4: connected component}
\label{\detokenize{demonstrations:model-4-connected-component}}
Each of the three models so far deals with all possible pairs of vertices.
The number of possible pairs of vertices is quadratic in the number of vertices.
However, many graphs are sparse, with a number of edges that is bounded by a linear function of the number of vertices.
For sparse graphs, and especially those with many vertices, it is therefore important to only consider the edges that are present rather than all possible pairs of vertices.
The next model \sphinxcode{\sphinxupquote{gc4.essence}} uses this insight, and is indeed faster than any of the three previous ones.

The model builds on the fact that a graph is disconnected if, and only if, its vertices can be partitioned into two sets, with no edges between vertices in the two different sets.
Here C is used to indicate a subset of the vertices.
There are three constraints.
The first is that C must contain some vertex.
The second is that C must be a connected component; each vertex in C is connected to some other vertex in C (unless C only contains a single vertex).
The third is that the value of \sphinxcode{\sphinxupquote{connected}} is determined by whether it is possible to find some vertex that is not in C.
The following is an attempt to capture these constraints in an Essence specification.

\begin{sphinxVerbatim}[commandchars=\\\{\}]
\PYG{k+kt}{given} n : \PYG{k+kt}{int}\PYG{o}{(}1..\PYG{o}{)}
\PYG{k+kt}{letting} vertices \PYG{k+kt}{be} \PYG{k+kt}{domain} \PYG{k+kt}{int}\PYG{o}{(}1..n\PYG{o}{)}
\PYG{k+kt}{given} G : \PYG{k+kt}{set} \PYG{k+kt}{of} \PYG{k+kt}{set} \PYG{o}{(}size 2\PYG{o}{)} \PYG{k+kt}{of} vertices
\PYG{k+kt}{find} C : \PYG{k+kt}{set} \PYG{k+kt}{of} vertices
\PYG{k+kt}{find} connected : bool
such that
  exists u : vertices . u in C,
  forAll e in G . \PYG{o}{(}min\PYG{o}{(}e\PYG{o}{)} in C\PYG{o}{)} = \PYG{o}{(}max\PYG{o}{(}e\PYG{o}{)} in C\PYG{o}{)},
  connected = !\PYG{o}{(}exists u : vertices . !\PYG{o}{(}u in C\PYG{o}{)}\PYG{o}{)}
\end{sphinxVerbatim}

This is the solution for \sphinxcode{\sphinxupquote{disconnected-4.param}}:

\begin{sphinxVerbatim}[commandchars=\\\{\}]
\PYG{k+kt}{letting} C \PYG{k+kt}{be} \PYG{o}{\PYGZob{}}1, 2\PYG{o}{\PYGZcb{}}
\PYG{k+kt}{letting} connected \PYG{k+kt}{be} false
\end{sphinxVerbatim}

Model \sphinxcode{\sphinxupquote{gc4.essence}} yields a solution quickly.
Unfortunately it can also give incorrect results: letting C be the set of all vertices and letting \sphinxcode{\sphinxupquote{connected}} be true is always a solution, whether the graph is connected or not.
This can be confirmed by asking Conjure to generate all solutions:

\begin{sphinxVerbatim}[commandchars=\\\{\}]
conjure solve \PYGZhy{}ac \PYGZhy{}\PYGZhy{}number\PYGZhy{}of\PYGZhy{}solutions=all gc4.essence
\end{sphinxVerbatim}

This gives two solutions, the one above and the following one:

\begin{sphinxVerbatim}[commandchars=\\\{\}]
\PYG{k+kt}{letting} C \PYG{k+kt}{be} \PYG{o}{\PYGZob{}}1, 2, 3, 4\PYG{o}{\PYGZcb{}}
\PYG{k+kt}{letting} connected \PYG{k+kt}{be} true
\end{sphinxVerbatim}

It is actually possible to ensure that this “solution” is never the first one generated, and then to ask Conjure to only look for the first solution; if the graph is not connected then the first solution will correctly indicate its status.
However, this relies on precise knowledge of the ordering heuristics being employed at each stage of the toolchain.

The problem with this fourth specification is that it only captures the property that C is a union of connected components.
We would need to add additional constraints to enforce the property that C should contain only one connected component.
This can be done, but is not especially efficent.

\subsection{Model 5: minimal connected component}
\label{\detokenize{demonstrations:model-5-minimal-connected-component}}
Let’s look for a robust approach that won’t unexpectedly fail if parts of the toolchain change which optimisations they perform or the order in which evaluations occur.

One option could be to look for solutions of a more restrictive model which includes an additional constraint that requires some vertex to not be in C.
This model would have a solution precisely if the graph is \sphinxstyleemphasis{not} connected.
Failure to find solutions to this model would then indicate connectivity.
It is possible to call Conjure from a script that uses the failure to find solutions to conclude connectivity, but the Conjure toolchain currently does not support testing for the presence of solutions directly.

In place of the missing “if-has-solution” directive, we could instead quantify over all possible subsets of vertices.
Such an approach quickly becomes infeasible as n grows (and is much worse than the models considered so far), because it attempts to check 2**n subsets.

As another option, we can make use of the optimisation features of Essence to find a solution with a C of minimal cardinality.
This ensures that C can only contain one connected component.
Choosing a minimal C ensures that when there is more than one solution, then the one that is generated always indicates the failure of connectivity.
Since we don’t care about the minimal C, as long as it is smaller than the set of all vertices if possible, we also replace the general requirement for non-emptiness by a constraint that always forces the set C to contain the vertex labelled 1.

\begin{sphinxVerbatim}[commandchars=\\\{\}]
\PYG{k+kt}{given} n : \PYG{k+kt}{int}\PYG{o}{(}1..\PYG{o}{)}
\PYG{k+kt}{letting} vertices \PYG{k+kt}{be} \PYG{k+kt}{domain} \PYG{k+kt}{int}\PYG{o}{(}1..n\PYG{o}{)}
\PYG{k+kt}{given} G : \PYG{k+kt}{set} \PYG{k+kt}{of} \PYG{k+kt}{set} \PYG{o}{(}size 2\PYG{o}{)} \PYG{k+kt}{of} vertices
\PYG{k+kt}{find} C : \PYG{k+kt}{set} \PYG{k+kt}{of} vertices
\PYG{k+kt}{find} connected : bool
such that
  1 in C,
  forAll e in G . \PYG{o}{(}min\PYG{o}{(}e\PYG{o}{)} in C\PYG{o}{)} = \PYG{o}{(}max\PYG{o}{(}e\PYG{o}{)} in C\PYG{o}{)}
minimising \PYG{o}{\textbar{}}C\textbar{}
\end{sphinxVerbatim}

This model \sphinxcode{\sphinxupquote{gc5.essence}} is still straightforward, even with the additional complication to rule out incorrect solutions.
Out of the correct models so far, this tends to generate the smallest input files for the back-end constraint or SAT solver, and also tends to be the fastest.

\subsection{Generating all connected graphs}
\label{\detokenize{demonstrations:generating-all-connected-graphs}}
We now have a fast model for graph connectivity.
Let’s modify it as \sphinxcode{\sphinxupquote{gce1.essence}}, hardcoding n to be 4 and asking the solver to find G as well as C.

\begin{sphinxVerbatim}[commandchars=\\\{\}]
\PYG{k+kt}{letting} n \PYG{k+kt}{be} 4
\PYG{k+kt}{letting} vertices \PYG{k+kt}{be} \PYG{k+kt}{domain} \PYG{k+kt}{int}\PYG{o}{(}1..n\PYG{o}{)}
\PYG{k+kt}{find} G : \PYG{k+kt}{set} \PYG{k+kt}{of} \PYG{k+kt}{set} \PYG{o}{(}size 2\PYG{o}{)} \PYG{k+kt}{of} vertices
\PYG{k+kt}{find} C : \PYG{k+kt}{set} \PYG{k+kt}{of} vertices
such that
  1 in C,
  forAll e in G . \PYG{o}{(}min\PYG{o}{(}e\PYG{o}{)} in C\PYG{o}{)} = \PYG{o}{(}max\PYG{o}{(}e\PYG{o}{)} in C\PYG{o}{)}
minimising \PYG{o}{\textbar{}}C\textbar{}
\end{sphinxVerbatim}

We now ask for all solutions:

\begin{sphinxVerbatim}[commandchars=\\\{\}]
conjure solve \PYGZhy{}ac \PYGZhy{}\PYGZhy{}number\PYGZhy{}of\PYGZhy{}solutions\PYG{o}{=}all gce1.essence
\end{sphinxVerbatim}

However, this finds only one solution!

The solver finds one solution that minimises \sphinxcode{\sphinxupquote{\textbar{}C\textbar{}}}; this minimisation is performed globally over all possible solutions.
This is what we intended when G was given, but is not what we want if our goal is to generate \sphinxstyleemphasis{all} connected graphs.
We want to minimise C for each choice of G, producing one solution for each G.
Currently there is no way to tell Conjure that minimisation should be restricted to the decision variable C.

Checking whether there is a nontrivial connected component seems to be the most efficient model for graph connectivity, but it doesn’t work in the setting of generating all connected graphs.
We therefore need to choose one of the other models to start with, say the iterated adjacency matrix representation.

We now use this model of connectivity to enumerate the labelled connected graphs over the vertices \sphinxcode{\sphinxupquote{\{1,2,3,4\}}}.
Previously we checked connectivity of a given graph G.
We now instead ask the solver to find G, specifying that it be connected.
We do this by asking for the same adjacency matrix \sphinxcode{\sphinxupquote{reach}} as before, but in addition asking for the graph G.
We also hardcode n, so no parameter file is needed, and add the condition that previously determined the value of the \sphinxcode{\sphinxupquote{connected}} decision variable as a constraint.

\begin{sphinxVerbatim}[commandchars=\\\{\}]
\PYG{k+kt}{letting} n \PYG{k+kt}{be} 4
\PYG{k+kt}{letting} vertices \PYG{k+kt}{be} \PYG{k+kt}{domain} \PYG{k+kt}{int}\PYG{o}{(}1..n\PYG{o}{)}
\PYG{k+kt}{find} G : \PYG{k+kt}{set} \PYG{k+kt}{of} \PYG{k+kt}{set} \PYG{o}{(}size 2\PYG{o}{)} \PYG{k+kt}{of} vertices
\PYG{k+kt}{letting} m \PYG{k+kt}{be} sum\PYG{o}{(}[1 \PYG{o}{\textbar{}} i : \PYG{k+kt}{int}\PYG{o}{(}0..64\PYG{o}{)}, 2**i \PYGZlt{}= n]\PYG{o}{)}
\PYG{k+kt}{find} reach : matrix indexed by [int\PYG{o}{(}0..m\PYG{o}{)}, vertices, vertices] \PYG{k+kt}{of} bool
such that
  forAll u,v : vertices . reach[0,u,v] = \PYG{o}{(}\PYG{o}{\PYGZob{}}u,v\PYG{o}{\PYGZcb{}} in G\PYG{o}{)},
  forAll i : \PYG{k+kt}{int}\PYG{o}{(}0..\PYG{o}{(}m\PYGZhy{}1\PYG{o}{)}\PYG{o}{)} . forAll u,v : vertices . reach[i+1,u,v] =
    \PYG{o}{(}reach[i,u,v] \PYGZbs{}/ \PYG{o}{(}exists w : vertices . \PYG{o}{(}reach[i,u,w] /\PYGZbs{} reach[i,w,v]\PYG{o}{)}\PYG{o}{)}\PYG{o}{)},
  forAll u,v : vertices . reach[m,u,v]
\end{sphinxVerbatim}

If this model is in the file \sphinxcode{\sphinxupquote{gce2.essence}}, then we now need to explicitly ask Conjure to generate all the possible graphs:

\begin{sphinxVerbatim}[commandchars=\\\{\}]
conjure solve \PYGZhy{}ac \PYGZhy{}\PYGZhy{}number\PYGZhy{}of\PYGZhy{}solutions\PYG{o}{=}all gce2.essence
\end{sphinxVerbatim}

In this case Conjure generates 38 solutions, one solution per file.

Instead of listing the edges of a graph, and then deriving the adjacency matrix as necessary, it is also possible to use the adjacency matrix representation.
As an exercise, modify the models of connectivity to use the adjacency matrix representation instead of the set of edges representation.
\phantomsection\label{\detokenize{zreferences:zreferences}}

\chapter{Contact}
\label{\detokenize{contact:contact}}\label{\detokenize{contact:id1}}\label{\detokenize{contact::doc}}
Conjure’s main developer is \sphinxhref{http://ozgur.host.cs.st-andrews.ac.uk}{Özgür Akgün}.
Please get in touch via \sphinxhref{mailto:ozgur.akgun@st-andrews.ac.uk}{email} if you have comments, suggestions, or if you encounter any problems.

You can also use the \sphinxhref{https://github.com/conjure-cp/conjure/issues}{issue tracker} to report bugs.

We are particularly interested in hearing specific comments about the documentation.
Please let us know if something is hard to understand, not easy to follow, or if the documentation is too sparse at a certain place.
We will do our best to help!

\section{Contributors}
\label{\detokenize{contact:contributors}}
The following list of people have contributed to the development of Conjure.
\begin{itemize}
\item {} 
\sphinxhref{http://ozgur.host.cs.st-andrews.ac.uk}{Özgür Akgün}

\item {} 
\sphinxhref{http://www-users.cs.york.ac.uk/frisch}{Alan Frisch}

\item {} 
\sphinxhref{http://ipg.host.cs.st-andrews.ac.uk}{Ian Gent}

\item {} 
\sphinxhref{http://homes.ieu.edu.tr/bhnich}{Brahim Hnich}

\item {} 
\sphinxhref{http://bh246.host.cs.st-andrews.ac.uk}{Bilal Syed Hussain}

\item {} 
\sphinxhref{http://caj21.host.cs.st-andrews.ac.uk}{Chris Jefferson}

\item {} 
\sphinxhref{http://ijm.host.cs.st-andrews.ac.uk}{Ian Miguel}

\item {} 
\sphinxhref{http://pwn1.host.cs.st-andrews.ac.uk}{Peter Nightingale}

\end{itemize}

\begin{sphinxthebibliography}{FHJ+08}
\bibitem[Dud24]{zreferences:dudeney1924puzzle}
H. E. Dudeney. Perplexities. \sphinxstyleemphasis{Strand Magazine}, 68:94,214, July 1924.
\bibitem[FHJ+08]{zreferences:frisch2008essence}
Alan M Frisch, Warwick Harvey, Chris Jefferson, Bernadette Martínez-Hernández, and Ian Miguel. Essence: a constraint language for specifying combinatorial problems. \sphinxstyleemphasis{Constraints}, 13(3):268\textendash{}306, 2008. \sphinxhref{https://doi.org/10.1007/s10601-008-9047-y}{doi:10.1007/s10601-008-9047-y}.
\end{sphinxthebibliography}

\renewcommand{\indexname}{Index}
\printindex
\end{document}